\documentclass{jkas}
\include{graphicsx}

\def\year{2022} 
\def\month{April} 
\def\volume{00} 
\def\issue{0} 
\def\beginpage{1} 
\def\endpage{99} 
\def\received{---} 
\def\accepted{---} 

\date{Received \received ; accepted \accepted}

\title{
Free-Floating Planets, the Einstein Desert, and 'Oumuamua
}

\author[1,2]{Andrew Gould}
\author[3,4]{Youn Kil Jung}
\author[3]{Kyu-Ha Hwang}
\author[5]{Subo Dong}
\author[6]{Michael D. Albrow}
\author[3]{Sun-Ju Chung}
\author[7]{Cheongho Han}
\author[3]{Yoon-Hyun Ryu}
\author[7]{In-Gu Shin}
\author[8]{Yossi Shvartzvald}
\author[9]{Hongjing Yang}
\author[10]{Jennifer C. Yee}
\author[9]{Weicheng Zang}
\author[3,11]{Sang-Mok Cha}
\author[3]{Dong-Jin Kim}
\author[3,4]{Seung-Lee Kim}
\author[3]{Chung-Uk Lee}
\author[3]{Dong-Joo Lee}
\author[3,11]{Yongseok Lee}
\author[3,4]{Byeong-Gon Park}
\author[2]{Richard W. Pogge}

\affil[1]{Max-Planck-Institute for Astronomy, K\"{o}nigstuhl 17, 69117 Heidelberg, Germany}
\affil[2]{Department of Astronomy Ohio State University,
140 W.\ 18th Ave., Columbus, OH 43210, USA 
\email{gould@astronomy.ohio-state.edu }}
\affil[3]{Korea Astronomy and Space Science Institute, Daejon
34055, Republic of Korea}
\affil[4]{Korea University of Science and Technology, Korea, 
(UST), 217 Gajeong-ro, Yuseong-gu, Daejeon, 34113, Republic of Korea}
\affil[5]{Kavli Institute for Astronomy and Astrophysics, Peking University,
Yi He Yuan Road 5, Hai Dian District, Beijing 100871, China}
\affil[6]{University of Canterbury, Department of Physics and
Astronomy, Private Bag 4800, Christchurch 8020, New Zealand}
\affil[7]{Department of Physics, Chungbuk National University,
Cheongju 28644, Republic of Korea}
\affil[8]{Department of Particle Physics and Astrophysics, 
Weizmann Institute of Science, Rehovot 76100, Israel}
\affil[9]{Department of Astronomy, Tsinghua University, Beijing 100084, China}
\affil[10]{Center for Astrophysics $|$ Harvard \& Smithsonian, 60 Garden
St., Cambridge, MA 02138, USA}
\affil[11]{School of Space Research, Kyung Hee University,
Yongin, Kyeonggi 17104, Republic of Korea}

\def\corrauthor{
Andrew Gould
}

\def\runningauthor{
Andrew Gould
}

\def\runningtitle{
FFPs, Einstein Desert, \& 'Oumuamua
}

\def\keywords{
gravitational microlensing, asteroids
}

\def\abstracttext{
We complete the survey for finite-source/point-lens (FSPL) giant-source
events in 2016-2019 KMTNet microlensing data.  The 30 FSPL events show
a clear gap in Einstein radius, $9\,\muas<\theta_\e <26\,\muas$, which is 
consistent with the gap in Einstein timescales near $t_\e\sim 0.5\,$days
found by \citet{mroz17} in an independent sample of
point-source/point-lens (PSPL) events.  We demonstrate that the two
surveys are consistent.  We estimate that the 4 events below this gap 
are due to a power-law distribution of free-floating planet candidates (FFPs)
$dN_{\rm FFP}/d\log M = (0.4\pm 0.2)\,(M/38 M_\oplus)^{-p}$/star, with 
$0.9\la p \la 1.2$.  There are substantially more FFPs than known
bound planets, implying that the bound planet power-law index $\gamma=0.6$ is
likely shaped by the ejection process at least as much as by formation.
The mass density per decade of FFPs in the Solar neighborhood is of the
same order as that of 'Oumuamua-like objects.  In particular, if we
assume that 'Oumuamua is part of the same process that ejected the
FFPs to very wide or unbound orbits, the
power-law index is $p=0.92\pm 0.06$.  If the Solar System's endowment
of Neptune-mass objects in Neptune-like orbits is typical, which is
consistent with the results of \citet{wide-orbit}, then these could account
for a substantial fraction of the FFPs in the Neptune-mass range.
}

\newcommand{\bdv}[1]{\mbox{\boldmath$#1$}}

\def\au{{\rm AU}}

\def\kms{{\rm km}\,{\rm s}^{-1}}
\def\masyr{{\rm mas}\,{\rm yr}^{-1}}
\def\muas{{\mu\rm as}}
\def\mas{{\rm mas}}
\def\kpc{{\rm kpc}}

\def\pc{{\rm pc}}

\def\rot{{\rm rot}}
\def\max{{\rm max}}
\def\rel{{\rm rel}}

\def\eff{{\rm eff}}

\def\hel{{\rm hel}}

\def\lim{{\rm lim}}
\def\e{{\rm E}}

\def\bmu{{\bdv\mu}}

\def\la{{\lesssim}}
\def\ga{{\gtrsim}}
\def\apj{{ApJ}}
\def\aj{{AJ}}
\def\apjl{{ApJL}}

\def\aap{{A\&A}}
\def\pasp{{PASP}}
\def\mnras{{MNRAS}}

\begin{document}
\jkashead 

\section{{Introduction}
\label{sec:intro}}

The claim by \citet{sumi11} of $\sim 2$ Jupiter-mass free-floating 
planet candidates (FFPs) per star
opened up the field of FFP population studies.
This led to several works that tried to interpret this
population \citep{clanton14,clanton14b,clanton16}.  However, it
was also later contradicted by \citet{mroz17}, who found no evidence for such
planets in a larger sample of events. Both of these
studies were limited to statistical analysis of the Einstein timescale
($t_\e$) distribution, for which the mass dependence is convolved with 
dependences on the lens-source relative proper motion ($\mu_\rel$)
and lens-source relative parallax ($\pi_\rel$).

\citet{kb192073} initiated a new approach to probing
the FFP population that focused on analyzing the Einstein radius 
($\theta_\e$) distribution. This approach has two advantages.
First, $\mu_\rel=\theta_\e/t_\e$ is automatically determined, which
removes one of the two convolutions with unknown distributions
that was mentioned above.  Second, the
selection function is relatively independent of lens mass ($M$) over a
broad range of masses.

When \citet{kb192073} reported KMT-2019-BLG-2073 as the fourth microlensing
single-lens event with a measured Einstein radius $\theta_\e<10\,\muas$, 
they noted that all four events had giant-star sources, which led them to
initiate a long-term study to find all finite-source/point-lens (FSPL)
events with such giant-star sources in the 2016-2019 Korea Microlensing
Telescope Network (KMTNet, \citealt{kmtnet}) database.  They
were motivated in part by the fact that this was the second giant-source
FSPL event from 2019 with $\theta_\e<10\,\muas$, the first being
OGLE-2019-BLG-0551 \citep{ob190551}.  They sought to place these FSPL
FFPs in the context of a homogeneously
selected sample that would include stars and brown dwarfs (BDs), as well
as FFPs.  For technical reasons that are briefly reviewed below, they
were only able to probe the 2019 database at that time.  Their systematic
search yielded a total of 13 giant-source FSPL events, including the two FFPs.
They carried out a variety of tests and concluded that there was nothing
suspicious about the sample.  However, they refrained from, and strongly
cautioned against, drawing any statistical conclusions from this sample,
despite the fact that it was homogeneously selected.  They argued that
because their
study was motivated by noticing two FFP detections in a single year,
it could suffer from publication bias.

Two factors prevented \citep{kb192073} from carrying out a 4-year search
immediately.  First, the form of the search had been made possible by
a recent upgrade to the 2019 online database, but this had not yet been
extended to the previous 3 years.  Second, for 2019, \citet{kb192073} 
had supplemented their automated analysis of the online database with
a special, more aggressive, execution of the EventFinder algorithm 
\citep{eventfinder} that was tailored to giant sources.  A more aggressive
search was required because the standard EventFinder algorithm takes 
advantage of the fact that most single-lens/single-source (1L1S) events
(and even most binary events) can be qualitatively matched to one of
two variants of \citet{gould96} 2-parameter profiles.  However, this is not
the case for typical FFP candidates, for which the source radius, $\theta_*$,
is often $\ga\theta_\e$, leading to a ``top hat'' light curve shape.
These upgrades and new searches required more than a year of effort.

The search proposed by \citet{kb192073} was therefore carried out as
these steps were completed season-by-season.  In the course of this work,
\citet{kb172820} reported the discovery of KMT-2017-BLG-2820, the 
third FFP in the KMTNet sample, which by this time was the sixth FSPL FFP
overall.  That is, in the intervening time, \citet{ob161928} had discovered
another FFP, OGLE-2016-BLG-1928.  \citet{kb172820} pointed out that
all six FFPs lay below an apparent ``gap'' in the cumulative 
$\theta_\e$ distribution from Figure~9 of \citet{kb192073}, which
they dubbed the ``Einstein Desert''.  

They argued from the definition of $\theta_\e$,
\begin{equation}
\theta_\e\equiv \sqrt{\kappa M \pi_\rel}\qquad
\kappa\equiv {4\, G\over c^2\au}\simeq 8.14\,{\mas\over M_\odot},
\label{eqn:thetaedef}
\end{equation}
that such a desert
would be a natural consequence of a bimodal function of the lens mass, $M$.
That is, the main stellar and BD lens population in the bulge (with 
small lens-source relative parallaxes, $\pi_\rel$) would contribute
relatively small Einstein radii, but would be cut off at 
$\theta_\e\sim 30\,\muas$ due to a sharp drop in the BD mass function, 
e.g., near $M\sim 0.02\,M_\odot$, combined with the paucity of phase
space for $\pi_\rel\la 5\,\muas$ (corresponding lens-source relative
distances $D_{LS}\equiv D_S - D_L \la 300\,\pc$).  Then, a second population 
of  low-mass FFPs, e.g., $M\sim 5\,M_\oplus$, with a much higher specific
frequency than stars and BDs would give rise to the small-$\theta_\e$ events.  

They then argued that a population of FFPs cannot be in the bulge and
produce the observed distribution, because an analogous population in
the disk would produce larger-$\theta_\e$ events that would ``fill in'' the
gap. By contrast, if the small-$\theta_\e$ events were explained by disk
FFPs, an analogous population in the bulge would not be
detected. Specifically, for
typical giant sources, the very numerous FFPs in the bulge would not
give rise to recognizable events because they would induce magnification, $A$,
changes of only 
\begin{equation}
A-1\simeq {2\over\rho^2} = {2\kappa M\pi_\rel\over\theta_*^2}= 0.13\,
\biggl({M \over 5\,M_\oplus}\biggr)
\biggl({\pi_\rel \over 20\,\muas}\biggr)
\biggl({\theta_* \over 6\,\muas}\biggr)^{-1}
\label{eqn:aminus1}
\end{equation}
where $\rho\equiv \theta_*/\theta_\e\ga 1$.  Hence, the FFP events would mainly
come from the less numerous disk population.  Then, a high inferred 
specific frequency would be necessary to compensate for the overall 
relative paucity of disk lenses.  (In Section~\ref{sec:constraints},
we will show that this reasoning is only partly correct.)

\citet{kb172820} recognized that with only 13 FSPL events, 
the evidence from \citet{kb192073} for a desert was relatively weak.
However, they adduced two additional arguments.  First, they noted that
the apparent gap was not (yet) contradicted by the ongoing search for
FSPL events in 2018 and (part of) 2017 data.  Second, they pointed 
to an analogous desert in the distribution of Einstein timescales,
\begin{equation}
t_\e\equiv {\theta_\e\over \mu_\rel}\qquad
\mu_\rel \equiv |\bmu_\rel|,
\label{eqn:tedef}
\end{equation}
found by \citet{mroz17} in their analysis of point-source/point-lens
(PSPL) events extracted from 5 years of
Optical Gravitational Lensing Experiment (OGLE) data.
Here, $\bmu_\rel$ is the lens-source  relative proper motion.
The same argument predicts a desert in the distribution of Einstein timescales, 
although it is expected to be less pronounced than the one in Einstein radii 
because the former is a convolution of the proper-motion distribution with 
the latter.

Thus, this picture gave a coherent account of all available data at that
time.  Nevertheless, \citet{kb172820} counseled a wait-and-see approach.
If their hypothesis were correct, they predicted that the Einstein
desert would remain relatively parched as new FSPL events were added.
If not, one should expect that it would be gradually ``filled in''.

Here, we present the full sample of 30 FSPL giant-source events from
the KMTNet 2016-2019 database.  We find that the Einstein Desert
is substantially more distinct than in the 13-event sample from 2019.
This strengthens the case for a bimodal mass function, with a broad
peak of stars and BDs complemented by a large population of FFPs
at much lower masses.  

We show that the FFPs are more numerous than the known bound planets.
We raise the possibility that the heavier FFPs have mainly been ejected
to very wide (rather than unbound) orbits, and may largely be accounted
for by the still poorly traced population of very-wide-orbit planets.
We also suggest that interstellar asteroids may be the extreme end of the
FFP population and, in particular, may follow the same power law.

\section{{Review of Detection Procedures}
\label{sec:detect}}

The FSPL events reported here are derived from the KMTNet microlensing survey,
which uses three identical 1.6m telescopes, each equipped with 
$2^\circ\times 2^\circ$ cameras \citep{kmtnet}.  The telescopes are located
in Chile (KMTC), South Africa (KMTS), and Australia (KMTA).  See Figure~12
from \citet{eventfinder} for the field locations and cadences.

We essentially follow the same procedures for selecting FSPL
preliminary candidates as was described by \citet{kb192073}.
We refer the reader to that paper for details.  Here we give
only a brief overview, and we also describe two important differences
between their 2019 sample and the 2016-2018 additional events that
we present here.  One of these was already mentioned by \citet{kb192073},
while the other is new.

We first conduct a purely automated search of the online summary table of
KMT events (about 3000 events per year), 
making use of the impact parameter $u_0$, the Einstein timescale
$t_\e$, and the source magnitude $I_S$ (from the pipeline
\citealt{pac86} PSPL fit), together with the
tabulated field extinction\footnote{The KMTNet webpage adopts:
$A_I = 7\,A_K$, where $A_K$ is from \citealt{gonzalez12}.}
$A_I$, to impose two selection conditions,
\begin{equation}
I_{S,0,\rm est} \equiv I_S - A_I < 16,
\label{eqn:giantstar}
\end{equation}
and
$$
\mu_{\rm thresh}\equiv {\theta_{*,\rm est}\over u_0 t_\e} > 1\,\masyr;
$$
\begin{equation}
\theta_{*,\rm est}\equiv 3\,\muas \times 10^{(16-(I_S - A_I))/5}.
\label{eqn:muthresh}
\end{equation}
The first condition restricts the sample to giants.  On rare occasions,
it allows foreground main-sequence or subgiant stars, but these are
eliminated at the stage of detailed investigations.

The second condition rejects the great majority of PSPL events while
accepting (for further investigation) the overwhelming majority of
FSPL events that satisfy Equation~(\ref{eqn:giantstar}).  First note
that the $\theta_{*,\rm est}$ estimate implicitly assumes that the source
has a similar color as the clump.  This will be approximately so except
for stars well up on the giant branch, for which it will be an underestimate.
Then, assuming that the online fits for PSPL events (i.e., $\rho<u_0$) are
approximately accurate, PSPL events will have
\begin{equation}
\mu_{\rm thresh}^{\rm pspl} = 0.33\,{\mas\over {\rm yr}}
\biggl({u_0\over 1/3}\biggr)^{-1}
\biggl({t_\e\over 20\,{\rm day}}\biggr)^{-1}
\biggl({\theta_{*,\rm est}\over 6\,\muas}\biggr).
\label{eqn:muthresh2}
\end{equation}
Hence, most PSPL giant-source events will fail this criterion, leaving
a tractable subset to be rejected by manual light-curve fitting.

On the other hand, for FSPL events, we do not expect the automated fit
to yield the correct $u_0$.  For $\rho \ll 1$, we expect 
$u_{0,\rm fit}t_{\e,\rm fit}\equiv t_{\eff,\rm fit} \simeq t_{*,\rm true}
\equiv\rho_{\rm true}t_{\e,\rm true}$ to the extent that the fit is influenced
by the width of the peak, and
$1/u_{0,\rm fit}\simeq A_{\rm max,\rm fit}\simeq A_{\rm max,\rm true} 
\simeq 2/\rho_{\rm true}$ to the extent that the fit is influenced
by the height of the peak.  These imply $\mu_\rel\sim \mu_{\rm thresh}$ 
and $\mu_\rel \sim \mu_{\rm thresh}/2$, respectively.  Hence,
FSPL events that fail Equation~(\ref{eqn:muthresh}) have relative proper
motions $\mu_\rel\la 1\,\masyr$.  Using the
approach in the Appendix to \citet{gould21}, the probability for a lens to
have such a small proper motion is 
$p<(\mu_{\rm thresh}/\sigma_\mu)^3/6\sqrt{\pi}\rightarrow 4\times 10^{-3}$,
where we have adopted an isotropic bulge proper-motion dispersion of
$\sigma_\mu\sim 2.9\,\masyr$.  Table~4 of \citet{kb192073}, which lists
both the estimator $u_{\rm thresh}$ and the true value $\mu_\rel$ for 13
events shows that the former is conservative in the sense that
$\mu_\rel/\mu_{\rm thresh}\la 1$.  Only two events 
(OGLE-2019-BLG-0551 and OGLE-2019-BLG-1143) violate this inequality,
and they do so only mildly, both with ratios 1.2.  Note that neither
of the two FFPs in this sample (OGLE-2019-BLG-0551, $\rho=4.5$;
KMT-2019-BLG-2073, $\rho=1.1$) are in the $\rho\ll 1$ regime, but both
approximately satisfy the inequality, with
$\mu_\rel/\mu_{\rm thresh}=1.2$ and 0.6, respectively.

Each such candidate, typically of order 60 per year, is first
reviewed visually.  Of order 10\% are eliminated at this stage
for various reasons, primarily that they are not single lens events
or that there are no data near the peak, which would be required to measure
$\rho$.  The remainder are fitted to PSPL and FSPL forms using pipeline
KMT data.  If FSPL is preferred by $\Delta\chi^2_{\rm fspl}>3$ and the best fit 
has a normalized impact parameter $z_0 \equiv u_0/\rho<1$ then the data are
subjected to tender loving care (TLC) re-reduction.  Then, if
$\Delta\chi^2_{\rm fspl}>20$, the event is selected as FSPL.  For a few
cases, the pipeline light curve has an obvious FSPL form for which
PSPL would give a very poor fit.  For these, we move directly from
inspection to TLC.  Note that both the pipeline reductions and TLC
reductions use pySIS \citep{albrow09}, which is a specific implementation
of difference image analysis \citep{tomaney96,alard98}.

For 2019, \citet{kb192073} separately searched the output of the
real-time event selection algorithm AlertFinder \citep{alertfinder}
and the end-of-year selection algorithm EventFinder \citep{eventfinder}.
Because AlertFinder did not operate in the wings of the season and
also because it searches only for rising events, there were wing-of-season
events that it missed but were found by EventFinder.  However, even in the
region of overlap, there were FSPL events that were missed by one
or the other.  For example the FFP FSPL event OGLE-2019-BLG-0551 was
found by AlertFinder but not EventFinder.  These discrepancies motivated
the addition of an aggressive giant-source search, which had two purposes.
First, it served as a check that the union of the AlertFinder and EventFinder
samples for 2019
was effectively complete.  And, indeed, the special giant-source search
did not yield any additional FSPL events.  However, \citet{kb192073}
also noted that 2019 was the first year for which AlertFinder was in full
operation.  For 2018, AlertFinder operated only beginning in June and only
in the northern bulge.  It did not operate at all in 2016 and 2017.
Hence, given that EventFinder was missing FSPL events in 2019, the special
giant-source search would have to be applied to previous seasons.  
See \citet{kb172820} for further refinements of the giant-source search.

Finally, we modified the search procedures relative \citet{kb192073}
to better deal with saturated
data.  In 2019, events that were saturated at peak were simply eliminated
as uninterpretable at the visual-inspection stage.  However, we subsequently
realized that many such events could be analyzed
by incorporating the $V$-band data for the
regions of the light curve that were saturated in the $I$ band.
For example, given the typical intrinsic colors of giants $(V-I)_0\sim 1.1$,
typical reddening of KMT fields $E(V-I)\sim 1.8$ and the roughly 0.65
deeper photometric zero point in the $V$ band, the onset of saturation
would be $\sim (1.1 + 1.8 - 0.65) = 2.25$ mag fainter in the $V$ band than
the $I$ band.  Of course, $V$-band data were taken 10 times less frequently
in 2017-2019 than $I$ band, and even less frequently in 2016.  However,
for events that are well characterized by bright (so, generally small-error)
$I$-band light curves, only a few points are needed over peak 
to measure $\rho$.  We applied this approach to the saturated events
from 2019, but this did not lead to new FSPL events.  However, it did prove
useful in fully characterizing FSPL events from 2016-2018.

\section{{Expected $\theta_\e$ Distribution}
\label{sec:expected}}

Our primary motivation is to probe the poorly understood population
of FFPs.  To do so, it is useful to understand the $\theta_\e$
distribution that is expected from stars and BDs.  We first show
that a substantial majority of these detections will come from bulge lenses
and second that these are expected to cover the range 
$30\,\muas\la\theta_\e\la 300\,\muas$.  We then turn to the role
of disk lenses.

Ignoring selection effects, the rate of microlensing events for
a population of fixed mass $M$ and toward a given source is
\begin{equation}
\Gamma_{\rm all}\propto \int \mu_\rel\theta_\e(D_L)D_L^2 d D_L \nu 
f(\mu_\rel)\mu_\rel g(M;D_L), 
\label{eqn:gammaall}
\end{equation}
where $\nu=\nu(D_L)$
is the density at distance $D_L$, $f(\mu) =
f(\mu_\rel;D_L)$ is the 1-dimensional proper motion distribution
and $g(M,D_L)$ is the mass function.  However, for FSPL events,
we replace $\theta_\e(D_L)\rightarrow \rho\theta_\e(D_L)=\theta_*$,
which eliminates one instance of implicit dependence on $D_L$ 
(and $M$; \citealt{gould13}).
We note that with the exception of very nearby lenses $D_L\la 1\,\kpc$
(see below), the mean proper motion $\langle\mu_\rel\rangle$ is
approximately independent of distance, while we also approximate 
$g(M)$ as being independent of distance.  Hence, for FSPL events,
\begin{equation}
\Gamma_{\rm FSPL}\propto \theta_*\langle\mu_\rel\rangle g(M)
\int_{D_L=0}^{D_S} d D_L D_L^2  \nu(D_L).
\label{eqn:gammafspl}
\end{equation}
That is, $\Gamma_{\rm FSPL}$ is proportional to the total column of lenses
in the observation cone between the observer and the bulge source.  Without
any detailed model, it is clear that this column is 
dominated by bulge lenses: otherwise, e.g., there would be a cloud of
disk clump stars trailing toward brighter magnitudes from the observed
bulge clump on the color-magnitude diagram (CMD).  Hence, bulge stars will
dominate most of the $\theta_\e$ distribution, while disk lenses will
dominate only in regions that are inaccessible to bulge lenses.

The bulge mass function is well populated over the range 
$0.02\,M_\odot\la M \la 1\,M_\odot$.  The upper limit is set by the fact that
the bulge is primarily an old population for which the stars that were born
$M\ga 1\,M_\odot$ have now mainly evolved to become remnants, mostly 
white dwarfs.
The lower limit is an approximation for the steeply falling mass function in the
BD regime.  The bulge has a depth of order $1\,\kpc$, implying that the typical
range of lens-source relative parallaxes is 
$0.005\,\mas\la\pi_\rel\la 0.02\,\mas $, corresponding to lens-source 
separations $0.35\,\kpc\la (D_S-D_L)\la 1.5\,\kpc$.  Above the upper limit,
the product of the disk and bulge density distribution declines sharply.
Below the lower limit, the amount of available phase space declines.
None of the four limits just described is a hard boundary, but 
together they indicate a well-populated range, for 
$\theta_\e=\sqrt{\kappa M\pi_\rel}$ of
$30\,\muas\la\theta_\e\la 400\,\muas$.

Then, we should consider whether these intrinsic boundaries are
impacted by additional detection effects.
For typical giant-star source radii $\theta_*\sim 6\,\muas$, these limits
correspond to $A_\max \sim 2/\rho\sim 10$ and 135, respectively.  The lower
limit is not near any selection boundary, but saturation
becomes an issue at the upper limit. For typical giant sources
$I_0\sim 14.5$ and extinctions $A_I\sim 2$, the upper boundary implies
$I_{\rm peak}\sim 11.2$, which is well into the range of KMT saturation,
whose onset is seeing (and hence observatory) dependent.
As we discussed in Section~\ref{sec:detect}, it is sometimes possible
to recover from saturation using $V$-band data.  Nevertheless, the onset
of saturation steepens the decline in the upper range of $\theta_\e$ 
detections, leading us to adopt $30\,\muas\la\theta_\e\la 300\,\muas$
as the expected range for bulge lenses.

If we were to adopt a similar mass function for disk lenses, then
they would populate a range that is translated upward by a factor
$\sqrt{\pi_{\rel,\rm disk}/\pi_{\rel,\rm bulge}}\sim 2$ in $\theta_\e$, 
where we have adopted $\pi_{\rel,\rm disk}\sim 60\,\muas$
and $\pi_{\rel,\rm bulge}\sim 15\,\muas$ as ``typical'' values for the
disk and bulge, respectively.  Hence, the disk lenses would appear
as ``sprinkled in'' (and not generally individually identifiable)
over the range $60\,\muas\la\theta_\e\la 400\,\muas$, and they
would only start to dominate for $\theta_\e\ga 400\,\muas$.
The fact that the disk mass function extends above $M\ga M_\odot$
augments this effect but does not qualitatively change it. As just
mentioned, this high $\theta_\e$ regime is suppressed by saturation
effects.  However, another way to express the relation of the disk
and bulge contributions is that at fixed $\theta_\e$, the disk
lenses have $\sim 4$ times lower mass than the bulge lenses.
Hence, at $\theta_\e\sim 400\,\muas$, they are near the peak
of the mass function $M\sim 0.3\,M_\odot$, whereas the bulge lens
mass function at $M\sim 1\,M_\odot$
is much lower.  Hence, we expect that, contrary to the
main part of the $\theta_\e$ distribution, the upper range will be a comparable
mix of disk and bulge lenses.

\begin{table*}
\begin{center}
\caption{\textsc{Microlens Parameters for FSPL giant-star events}}
\begin{tabular}{llrrrrr}
\hline
\hline
\multicolumn{1}{c}{Name} &
\multicolumn{1}{c}{KMT Name} &
\multicolumn{1}{c}{$t_0$} &
\multicolumn{1}{c}{$u_0$} &
\multicolumn{1}{c}{$t_\e$} &
\multicolumn{1}{c}{$\rho$} &
\multicolumn{1}{c}{$f_{s,\rm KMTC}$} \\
\hline
OB180705 & KB181882 & 8316.30592 &  0.00123 & 46.247 & 0.01151 & 5.5354 \\
         & (errors) &    0.00145 &  0.01239 &  0.147 & 0.00503 & 0.0187 \\
KB180244 & KB180244 & 8312.42231 &  0.27149 &  4.430 & 0.35487 &11.0566 \\
         & (errors) &    0.00118 &  0.00301 &  0.027 & 0.00283 & 0.1300 \\
OB180626 & KB182309 & 8230.28800 &  0.16989 &  2.983 & 0.21484 &10.3031 \\
         & (errors) &    0.00132 &  0.00255 &  0.024 & 0.00251 & 0.1520 \\
MB17147  & KB170132 & 7850.99388 &  0.09138 &  2.695 & 0.13451 & 4.3062 \\
         & (errors) &    0.00043 &  0.00097 &  0.016 & 0.00102 & 0.0393 \\
OB171254 & KB170374 & 7952.25224 &  0.00000 & 15.278 & 0.02527 & 0.7068 \\
         & (errors) &    0.00025 &  0.00000 &  0.046 & 0.00009 & 0.0024 \\
OB170560 & KB172830 & 7859.52120 &  0.02900 &  0.901 & 0.89000 &18.5140 \\
         & (errors) &    0.00100 &  0.01800 &  0.002 & 0.00300 & 0.0010 \\
OB170905 & KB171022 & 7895.73816 &  0.08702 &  7.734 & 0.15031 &83.4744 \\
         & (errors) &    0.00401 &  0.00221 &  0.046 & 0.00254 & 0.9971 \\
MB17241  & KB170818 & 7883.47648 &  0.21697 &  1.845 & 0.29496 & 4.8304 \\
         & (errors) &    0.00057 &  0.00437 &  0.024 & 0.00483 & 0.1019 \\
OB170084 & KB170726 & 7807.13435 &  0.01226 & 43.642 & 0.02335 & 0.4049 \\
         & (errors) &    0.00212 &  0.00022 &  0.501 & 0.00028 & 0.0055 \\
OB161045 & KB160848 & 7559.20148 &  0.01061 & 12.030 & 0.02943 & 1.3546 \\
         & (errors) &    0.00115 &  0.00208 &  0.099 & 0.00191 & 0.0141 \\
KB161128 & KB161128 & 7486.32769 &  0.00951 & 12.550 & 0.02586 & 0.2703 \\
         & (errors) &    0.00147 &  0.00105 &  0.428 & 0.00105 & 0.0106 \\
OB161540 & KB162262 & 7606.72400 &  0.60500 &  0.330 & 1.63100 &19.6960 \\
         & (errors) &    0.00100 &  0.02700 &  0.003 & 0.00800 & 0.0000 \\
KB162057 & KB162057 & 7467.93171 &  0.00000 & 11.374 & 0.06645 & 2.4595 \\
         & (errors) &    0.00196 &  0.00000 &  0.105 & 0.00078 & 0.0309 \\
MB16258  & KB160606 & 7537.29800 &  0.18800 &  3.722 & 0.57400 &49.7270 \\
         & (errors) &    0.00100 &  0.00400 &  0.045 & 0.00300 & 0.4670 \\
\hline
\end{tabular}
 \tabnote{Notes: The units of $t_0$ and $t_\e$ are HJD$^\prime={\rm HJD}-2450000$
and days, respectively. Fluxes are in units of an $I=18$ system.
Event names are abbreviations for,
  e.g., OGLE-2018-BLG-0705.
 For KB172820, see \citet{kb172820}. 
 For OB170896 (=KB170799), see \citet{ob170896}.
 For OB171186 (=KB170357), see \citet{ob171186}.
 For OB180705, the fit included parallax:
 $(\pi_{\e,N},\pi_{\e,E})=
 (-0.324,-0.017)\pm (0.139,0.008)$.}
 \label{tab:parms}
\end{center}
\end{table*}

\begin{table*}
\begin{center}
\caption{\textsc{CMD and Derived Parameters for FSPL giant-star events}}
\begin{tabular}{llrrrrrrr}
\multicolumn{1}{c}{Name} &
\multicolumn{1}{c}{KMT Name} &
\multicolumn{1}{c}{$(V-I)_0$} &
\multicolumn{1}{c}{$I_0$} &
\multicolumn{1}{c}{$\theta_*$} &
\multicolumn{1}{c}{$\theta_\e$} &
\multicolumn{1}{c}{$\mu_\rel$} &
\multicolumn{1}{c}{$\mu_{\rm thresh}$} &
\multicolumn{1}{c}{$z_0=u_0/\rho$} \\
\hline
\hline
OB180705 & KB181882 &   1.46 & 13.14 &14.78 & 1285.00 & 10.10 & 13.31 & 0.104 \\
KB180244 & KB180244 &   2.26 & 12.06 &34.25 &   96.72 &  7.97 &  7.79 & 0.764 \\
OB180626 & KB182309 &   1.20 & 14.69 & 6.53 &   29.00 &  3.55 &  5.63 & 0.791 \\
MB17147  & KB170132 &   0.94 & 14.18 & 6.22 &   46.24 &  6.26 & 13.75 & 0.679 \\
KB172820 & KB172820 &   1.13 & 14.31 & 7.05 &    5.94 &  7.95 & 10.34 & 0.150 \\
OB171254 & KB170374 &   1.27 & 14.97 & 5.68 &  224.11 &  5.35 & 10.14 & 0.000 \\
OB170896 & KB170799 &   1.26 & 14.95 & 5.71 &  139.60 &  3.42 &  5.94 & 0.115 \\
OB170560 & KB172830 &   2.43 & 12.47 &31.10 &   34.90 & 14.17 &  5.80 & 0.033 \\
OB170905 & KB171022 &   1.60 & 11.19 &38.50 &  256.15 & 12.09 &  5.49 & 0.579 \\
MB17241  & KB170818 &   1.24 & 14.29 & 7.65 &   25.93 &  5.13 &  5.09 & 0.736 \\
OB171186 & KB170357 &   2.12 & 12.39 &26.90 &   93.90 &  2.65 &  2.88 & 0.319 \\
OB170084 & KB170726 &   1.33 & 15.41 & 4.78 &  204.11 &  1.71 &  2.38 & 0.525 \\
OB161045 & KB160848 &   1.23 & 14.37 & 7.33 &  249.31 &  7.55 & 43.72 & 0.360 \\
KB161128 & KB161128 &   1.00 & 14.43 & 5.49 &  212.79 &  6.19 & 24.10 & 0.368 \\
OB161540 & KB162262 &   1.70 & 13.58 &14.36 &    8.80 &  9.74 &  8.80 & 0.371 \\
KB162057 & KB162057 &   1.20 & 14.58 & 7.29 &  109.71 &  3.52 &  8.41 & 0.000 \\
MB16258  & KB160606 &   2.92 & 12.31 &43.89 &   76.45 &  7.50 &  1.65 & 0.328 \\
\hline
\end{tabular}
 \tabnote{Notes: The units of $\theta_*$ and $\theta_\e$ are $\muas$, while
those of $\mu_\rel$ and $\mu_{\rm thresh}$ are $\masyr$.
Event names are abbreviations for,
  e.g., KMT-2018-BLG-1882, OGLE-2018-BLG-0705, and 
  MOA-2017-BLG-147}
 \label{tab:fspl}
\end{center}
\end{table*}

We note that the higher proper motions of nearby lenses $D_L\la 1\,\kpc$
do not compensate for their low observing-cone volume in the rate equation.
The mean proper-motion term in this regime scales 
$\langle \mu_\rel(D_L) \rangle \sim v_{\perp,\rm typ}/D_L$ where 
$v_{\perp,\rm typ}\sim 40\,\kms \sim 8\,\kpc\,\masyr$ is the typical transverse
velocity of nearby stars.  Hence, when 
the proper-motion term is moved inside the integral, it takes the form
$v_{\perp,\rm typ}\int d D_L\,D_L \ldots$ in place of
$\langle\mu_\rel\rangle\int d D_L\,D_L^2 \ldots$ at larger distances.
That is, the nearby lens contribution is still heavily suppressed by
the volume factor.

Finally, it is reasonable to expect sensitivity to FSPL events with peak 
magnifications
down to $A_\max\sim 2$, i.e., $\rho\sim 1.4$ (Equation~(\ref{eqn:aminus1}), 
hence $\theta_\e\sim \theta_*/1.4 \sim 4\,\muas$.  Thus, the survey should
have sensitivity that goes almost a decade below the Einstein radii that are
generated by known stars and BDs.

\section{{30 FSPL Events from 2016-2019}
\label{sec:33events}}

Table~\ref{tab:parms} shows the microlens fit parameters for the 14 of
the 17 new FSPL events that were discovered in 2016-2018 data.  
The parameters for the remaining three are adopted from previous
publications.  See Appendix~\ref{sec:individual}.
For the 13 FSPL events from 2019, see Table~3 of \citet{kb192073}.  

For the color-magnitude analysis, we used the same procedures as 
\citet{kb192073}, ultimately based on \citet{ob03262}.  Briefly, we
perform a special reduction using pyDIA \citep{pydia}, which carries
out light-curve photometry and field-star photometry on the same
system.  We find $I_S$ by regression of the $I$-band light curve on
the best model and the $(V-I)_S$ color by regression of the $V$-band
on the $I$-band light curve.  We find the clump centroid of the field
stars, and so calculate the offset $\Delta[(V-I),I]$ of the source 
relative to the clump.  We estimate the intrinsic clump color
$(V-I)_{\rm cl,0}=1.06$ from \citet{bensby13} and its intrinsic magnitude
$I_{\rm cl,0}$ from Table~1 of \citet{nataf13}.  We convert from $V/I$
to $V/K$ using the color-color relations of \citet{bb88}.  Finally,
we estimate $\theta_*$ using the color/surface-brightness relations
of \citet{kervella04} for K giants and \cite{groenewegen04} for M giants.
The results of the CMD analysis are shown in Table~\ref{tab:fspl} 
for the 17 events
from 2016-2018.  For the 13 FSPL events from 2019, see Table~4 of 
\citet{kb192073}.  

Among the 17 newly reported FSPL events, we recovered the FFP 
OGLE-2016-BLG-1540 \citep{ob121323} (and, as already reported by
\citealt{kb172820}, the FFP KMT-2017-BLG-2820), but did not find any 
additional FFPs. Hence, there are a total of four FFPs in the sample 
of 30 FSPL events, OGLE-2016-BLG-1540, KMT-2017-BLG-2820, OGLE-2019-BLG-0551, 
and KMT-2019-BLG-2073.  

We did not recover either of the other two
known FSPL FFPs, OGLE-2012-BLG-1323 \citep{ob121323} 
and OGLE-2016-BLG-1928 \citep{ob161928}.  The first preceded the KMT survey.
For the second, the event was not found either by EventFinder nor in the
special giant-source search because there are only 4 magnified KMT points.
Even if identified, it could not have been reliably characterized from
these four points.  OGLE's original identification of this event was
post-season and was based on 10 magnified points that probe both the
peak and the wings.  The role of KMT data (in addition to confirming the
event from KMTC) was to rule out binary models by the flat behavior of
KMTS data starting about 8 hours after the last OGLE point.

In Appendix~\ref{sec:individual}, we provide notes on individual events.

\begin{figure}
\centering
\includegraphics[trim=0mm 10mm 20mm 10mm, width=90mm]{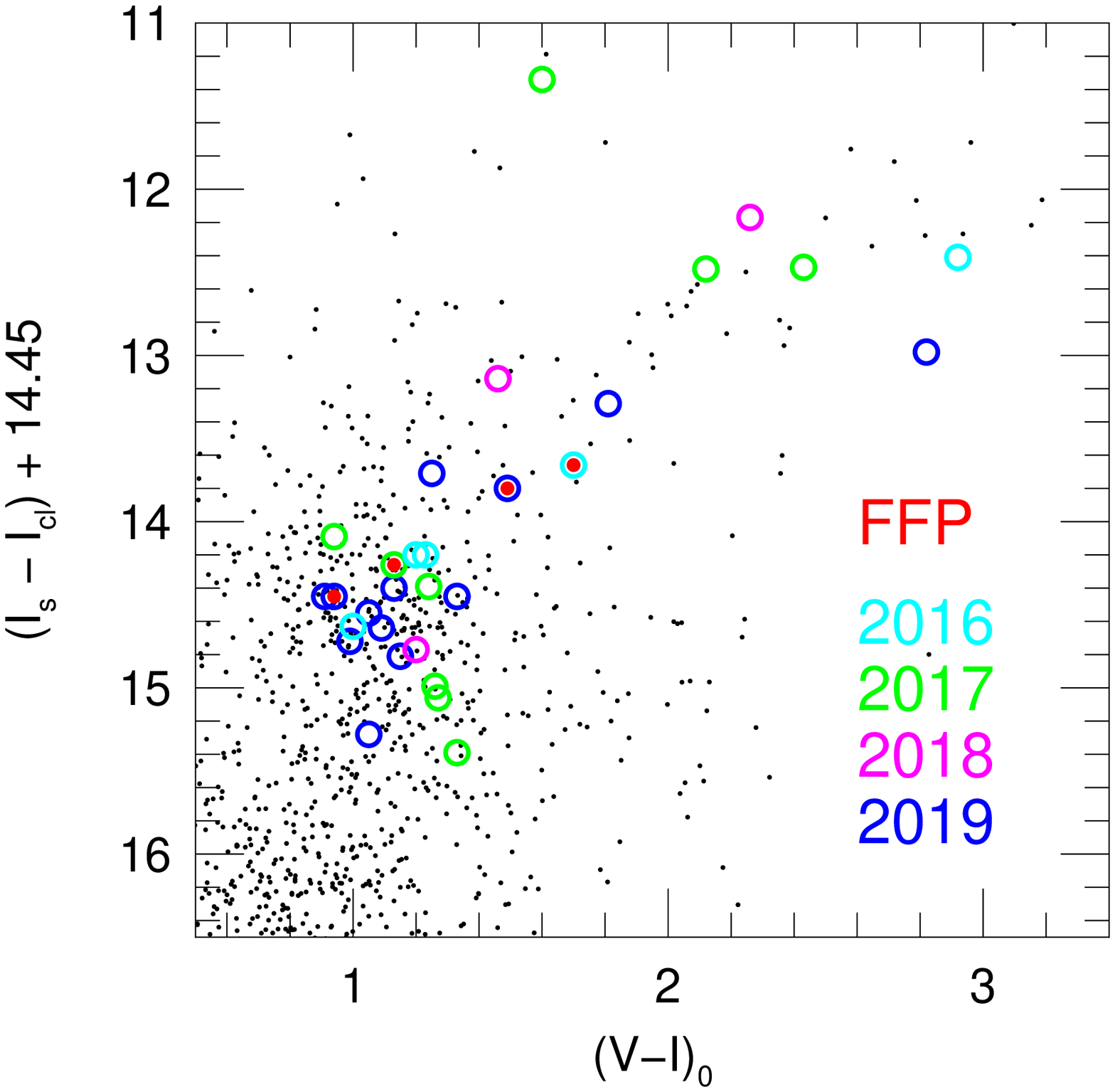}
\caption{Color-magnitude locations of the 30 FSPL events relative to
the clump, shown against a background of field stars from the
CMD of KMT-2019-BLG-2555, i.e., the same as was used in Figure~8 of 
\citet{kb192073}.  The events are color-coded by discovery year.
FFPs have red cores.  As expected, the sources are concentrated in
the clump (the highest concentration of giants), are somewhat 
over-represented on the upper giant branch (due to larger cross sections),
and, conversely somewhat under-represented on the lower giant branch.
The largest outlier, OGLE-2017-BLG-0905 (green point near top), is discussed in 
Appendix~\ref{sec:individual}, while the second largest, KMT-2019-BLG-1143 
 (blue point toward
the right), was discussed by \citet{kb192073}.  The lack of detections
with ordinate $\ga 15.5$ is discussed in Section~\ref{sec:sample}, with
reference to \citet{kb192073}.
}
\label{fig:cmdall}

\centering
\includegraphics[trim=0mm 10mm 20mm 30mm, width=90mm]{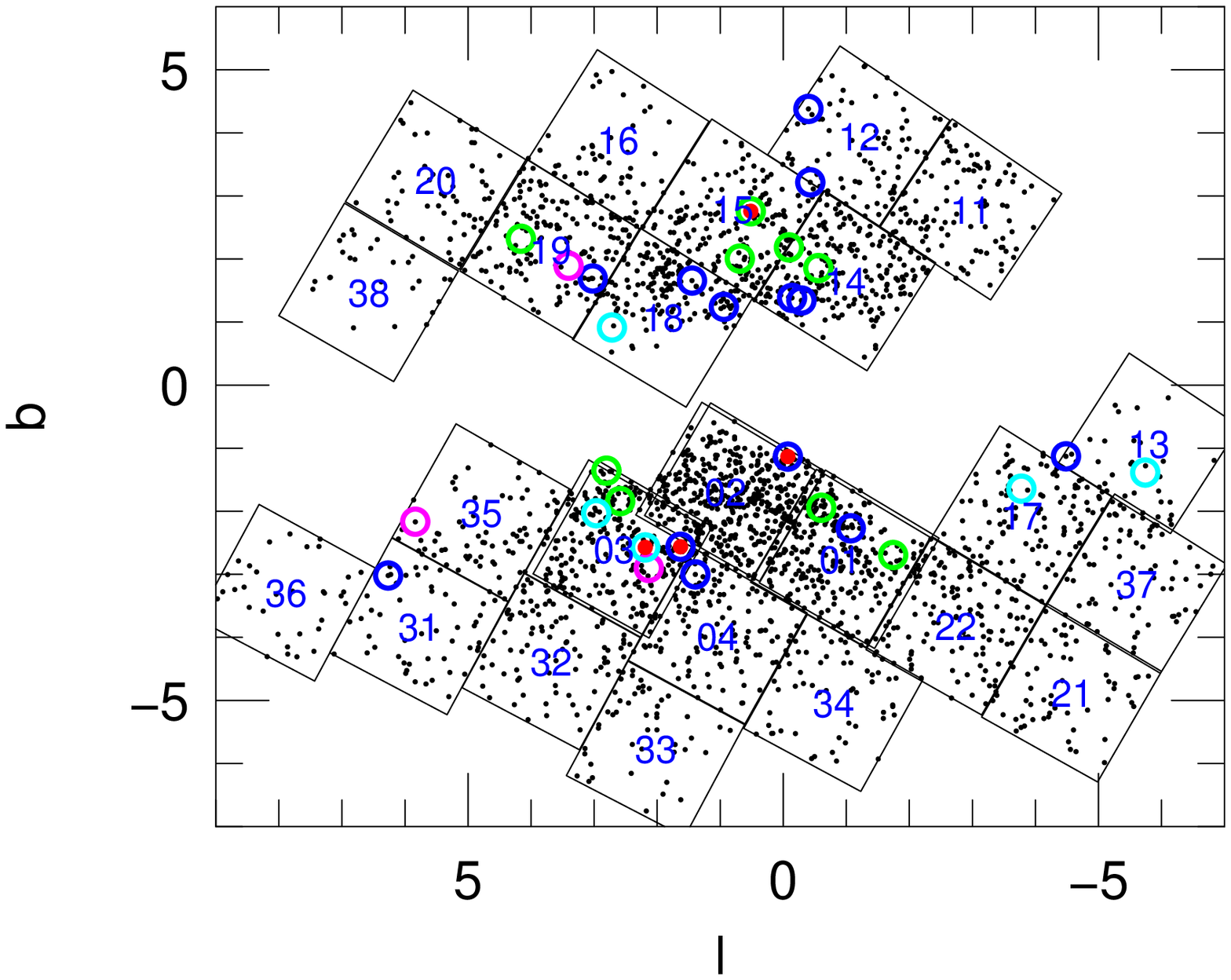}
\caption{Field positions of 30 FSPL events in Galactic coordinates, color-coded
accorded to year of discovery (see labels in Figure~\ref{fig:cmdall}).
The background black points are (as in Figure~4 of \citealt{kb192073})
EventFinder events from 2019.  As in that Figure, the KMT fields are
shown as black squares, with blue field numbers.  
Their observational cadences can be obtained from Figure~12 of
\citet{eventfinder}.
}
\label{fig:lb}
\end{figure}

\begin{figure}[t]
\centering
\includegraphics[trim=0mm 10mm 20mm 10mm, width=90mm]{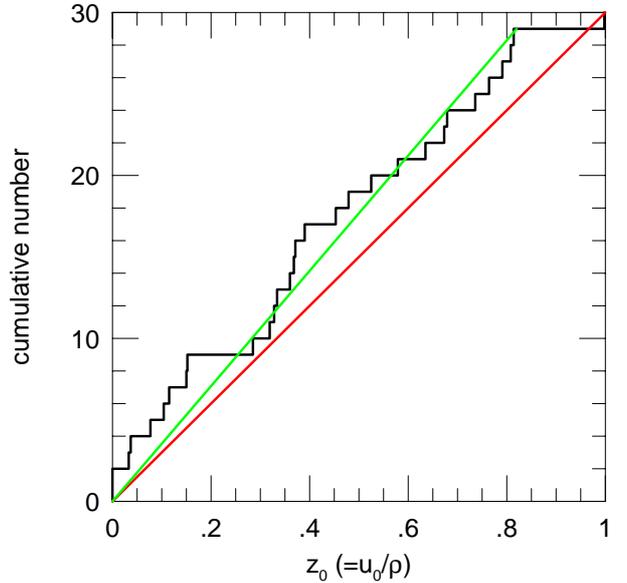}
\caption{Cumulative distribution of normalized impact parameters 
$z_0=u_0/\rho$.  The red line connects the first and last points,
while the green line connects the first and penultimate points.
The underlying distribution is rigorously uniform in $z_0$, so the
eye is struck by the absence of detections toward $z_0\simeq 1$.
According to a Kolmogorov-Smirnov (KS) test, this is not significant, 
but the effect
may be real because the KS test does not capture all the physics.
See Section~\ref{sec:sample}.  The distribution relative to the
green line suggests that the sample is homogeneously selected for
$z_0\la 0.8$.
}
\label{fig:z}
\end{figure}

\begin{figure}[t]
\centering
\includegraphics[trim=0mm 10mm 20mm 20mm, width=90mm]{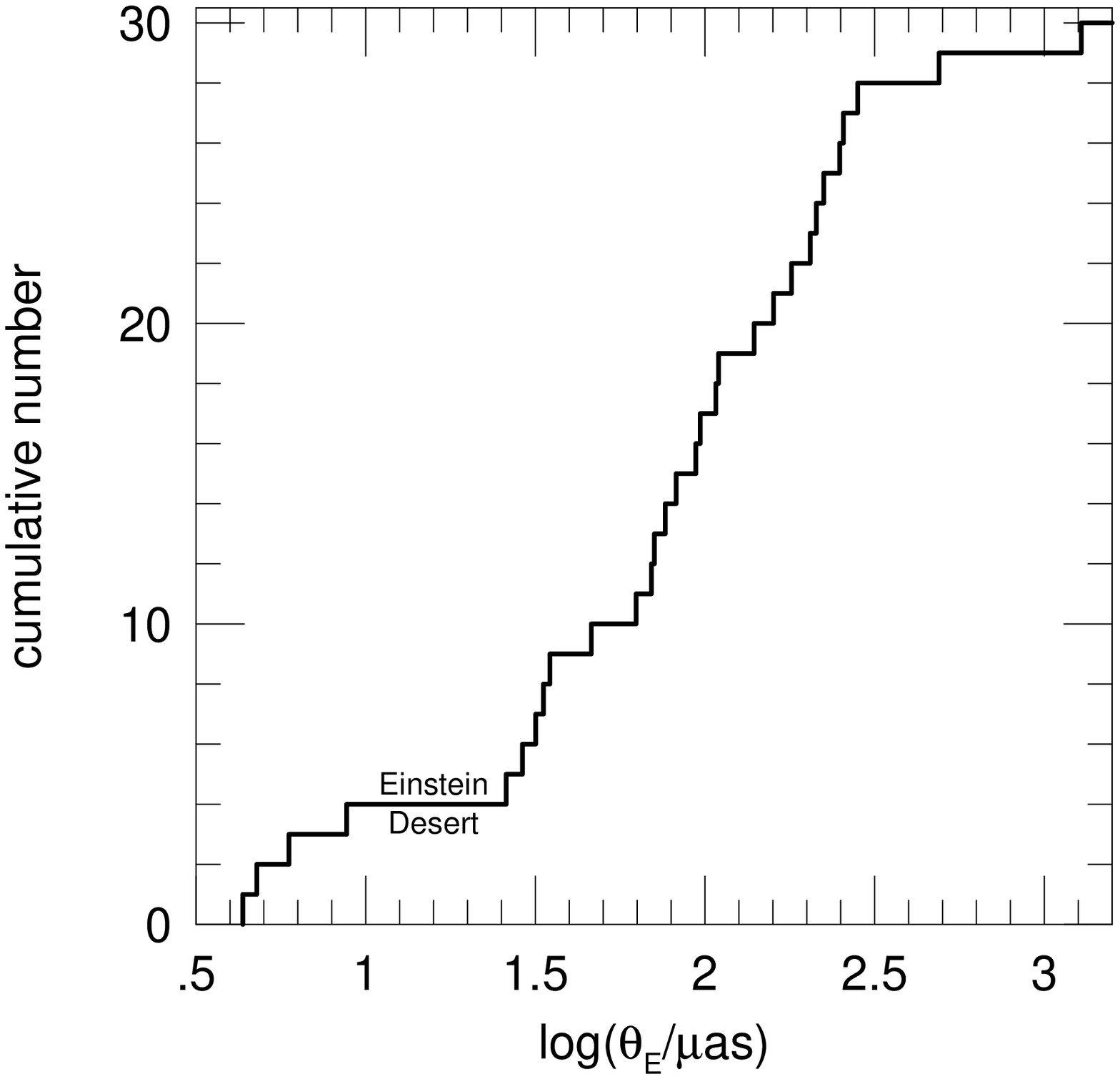}
\caption{Cumulative Distribution of $\log\theta_\e$ for
30 FSPL events with giant-star sources
that were found in four seasons of KMTNet data, 2016-2019.  The absence
of detections $\theta_\e\la 4\,\muas$ is due to selection effects, 
while the paucity for $\theta_\e\ga 300\,\muas$ is due to a combination
of selection effects and
the lack of bulge stars $M\ga 1\,M_\odot$.  
The remaining feature, the  factor $\sim 3$ ``Einstein Desert'',
$8.8\,\muas <\theta_\e< 26\,\muas$, is the signature of two populations, i.e.,
BDs/stars with masses $0.02\la M/M_\odot \la 1$ to the right and
FFPs with much lower masses to the left.
The former are detected mainly in the bulge, while the latter are
likely to be about equally drawn from the bulge and the disk.  
See Section~\ref{sec:power-law}.
This Figure can be compared to Figure~10 from \citet{kb192073}.
}
\label{fig:cum3}
\end{figure}

\section{{Sample Characteristics}
\label{sec:sample}}

Notwithstanding the justified caution of \citet{kb192073}, the results from 2019
prove to be reasonably representative of the four year sample.  For the four
years 2016-2019, the FFP/FSPL counts are (1/5, 1/9, 0/3, 2/13).  For
2016-2018, they are 2/17 compared to 2/13 for 2019, which is consistent
with normal Poisson variations.  If four detections are distributed
randomly among four seasons, the most common configuration (144/256)
is the observed one, (2,1,1,0).  The one feature that is possibly unusual
is that there were only 3 FSPL events in 2018 compared to 13 in 2019, so
one may wonder whether the seasonal distribution of FSPL
events, $n_i=(5,9,3,13)$ can also be considered as a ``typical''
outcome of Poisson sampling.  We investigate this by adopting
an annual expectation\footnote{Inspection of the formula for $\ln L$ shows
that the choice of $\lambda$ plays no role: it just adds a constant
$30\ln\lambda - 4\lambda\rightarrow 30.45$ to each trial.  That is,
we could have just as well used $\ln L = -\sum_{i=1}^4 \ln n_i !$.  We keep
the more complex form to maintain familiarity for the reader.}
$\lambda=30/4=7.5$ and construct a log likelihood 
$\ln L = \sum_{i=1}^4 \ln (e^{-\lambda}\lambda^{n_i}/n_i!) = -10.31 $.  We can then
compare this to the same statistic for $10^6$ realizations in which 30 events
are randomly distributed among 4 seasons.  We find that 7.8\% have smaller
likelihood.  Thus, this outcome is not particularly ``unexpected'', especially
given that this is a posterior test.
Therefore, neither the seasonal distribution of the FFPs nor that of
the FSPL events can be considered unusual.

\citet{kb192073} presented several diagnostic plots whose purpose
was to probe for statistical artifacts and/or patterns
in the sample and to identify
individual outliers that might require explanation and/or further
investigation.  We present analogous plots here.

Figure~\ref{fig:cmdall} shows the color-magnitude diagram (CMD)
source positions of the FSPL
events relative to the clump, color-coded by season.  The blue (2019) points
and the black background points (taken from the dereddened 
CMD of KMT-2019-BLG-2555) are the same as those shown in Figure~8 of 
\citet{kb192073}.  Of the 30 sources, 15 are clump stars (or giant-branch
stars that are superposed on the clump), 4 are lower-giant-branch
stars, 9 track the upper giant branch quite closely, 
one (KMT-2019-BLG-1143, blue) lies significantly below the upper giant branch,
and one (OGLE-2017-BLG-0905, green) lies well above the upper giant branch.

The broad pattern of the FSPL events relative to the background
is as expected: they are concentrated in the clump (i.e., the most
densely populated region of giant stars), and are somewhat over-represented
on the upper giant branch and under-represented on the lower giant branch,
which is expected due to larger and smaller cross sections, respectively.
The complete absence of FSPL events between $16<I<15.5$ would appear
to be somewhat surprising.  However, \citet{kb192073} showed that this
is due to systematic underestimation (with considerable scatter)
of source flux for FSPL events by the automated pipeline.  See their Figure~7.

The outlier KMT-2019-BLG-1143 was investigated by \citet{kb192073}.
We investigate the other outlier (OGLE-2017-BLG-0905)
in Appendix~\ref{sec:individual}.

Figure~\ref{fig:lb} shows the distribution of the 30 FSPL events on the
sky in Galactic coordinates, against the same background of 2019 
EventFinder events that appears in Figure~4 of \citet{kb192073}.
The FSPL events are concentrated toward the Galactic plane,
as would be expected for giant-source events because they
are more heavily represented in regions of high extinction.
However, this is a weak effect, and no strong statistical
conclusion can be drawn from this apparent pattern.

Figure~\ref{fig:z} shows the cumulative distribution of the 30 FSPL events
as a function of normalized impact parameter $z_0\equiv u_0/\rho$.  In nature,
this distribution is exactly uniform, so any (statistically significant)
deviation from a uniform distribution
should reflect selection effects.  \citet{kb192073}
had argued based on their analogous Figure~3, that the distribution was
consistent with being uniform over $[0\leq z_0 < 1]$.  Comparison of the
observed distribution with a uniform expectation (red line) indicates that,
with accumulating statistics, this may no longer be true.  
A Kolmogorov-Smirnov (KS) test yields an unimpressive $p=28\%$.  However,
KS is a very ``forgiving'' test because it assumes no prior information
on the form of the deviation.  By contrast, our eye is drawn to the
near absence of events with $z_0>0.82$, where we might expect the
greatest difficulty for detections.  The green line, which ignores the
single detection at $z_0\simeq 1$, appears more satisfying.
\cite{kb192073} argued that there is substantial information about $\rho$
in the regions of the light curve having $z\equiv u/\rho>1$, so that uniform
sensitivity over $[0\leq z_0 < 1]$ was plausible.  However, it is also
plausible that there would be at least some effect for $z_0\sim 1$, as
seems to be indicated by Figure~\ref{fig:z}, albeit weakly.

One might suspect that the dearth of events with $z_0\sim 0.9$ is the
result of the $z_0<1$ boundary in our initial selection combined
with ordinary statistical noise.  That is, some events with initial
estimates $z_{0,\rm init}\sim 0.9$ were ultimately eliminated after
they were determined to have $z_{0,\rm TLC}>1$ from subsequent 
TLC reductions, but the
``complementary'' events that would have had $z_{0,\rm TLC}\sim 0.9$ (if
TLC reductions had been done) were not investigated after it was found
that $z_{0,\rm init}> 1$.  In fact, we were concerned about this and obtained
TLC for 5 events with $z_{0,\rm init}\ga 1$.  However, $z_{0,\rm TLC}>1$ was 
confirmed for all 5.  Hence, we do not believe that such selection bias
is a strong effect.  

\section{{The Einstein Desert}
\label{sec:desert}}

Figure~\ref{fig:cum3} shows the cumulative distribution $\log\theta_\e$
for the 30 FSPL events in the survey.  It exhibits three principal features;
1: a paucity of detections with $\theta_\e\ga 300\,\muas$,
2: a complete absence of detections with $\theta_\e< 4\,\muas$, and
3: the ``Einstein Desert'', i.e., the absence of detections  over the interval, 
$8.8\,\muas< \theta_\e < 26\,\theta_\e$.

As discussed in Section~\ref{sec:expected}, the first was expected,
primarily due to the rapid fall-off of the bulge mass function for 
$M\ga 1\,M_\odot$, but somewhat augmented by selection effects due to 
saturation.  The latter will be examined more closely in 
Section~\ref{sec:select}.

As will also be discussed in Section~\ref{sec:select}, the second feature
is likewise due to selection.

However, selection effects play no role in the third feature,
the ``Einstein Desert'', because
the selection function is smoothly increasing from a factor of 2 below the
Desert to a factor of a few above it.

As was already anticipated in Section~\ref{sec:expected}, the upper shore of 
this desert (at $\theta_\e=26\,\muas$) can be attributed to the sharp fall-off
in the bulge mass function in the BD regime.  In this sense, it is qualitatively
similar to the first feature, but with a crucial difference.  Being generated
by a fall-off at the high end of the mass function, the first feature
is ``washed out'' by contributions of disk lenses of much lower mass
(hence, higher specific frequency) but with 
similar $\theta_\e=\sqrt{\kappa M\pi_\rel}$.  By contrast, the population
of disk stars and BDs only contribute to $\theta_\e$ in regions well above
the desert, and so the upper edge of the Desert is sharp.

The 4 events that lie below the Desert must represent a separate, low-mass
population because they have Einstein radii that are a factor 
$\ga 3$ below those of the lowest-mass bulge BD lenses (and even farther
below  those of the lowest-mass disk BD lenses).  The existence of the Desert
is a powerful constraint on the nature of this population: whatever model
one adopts must not only reproduce the observed low-$\theta_\e$ detections
but must also respect the absence of such detections in the Desert.

\begin{figure}
\centering
\includegraphics[trim=0mm 10mm 20mm 20mm, width=90mm]{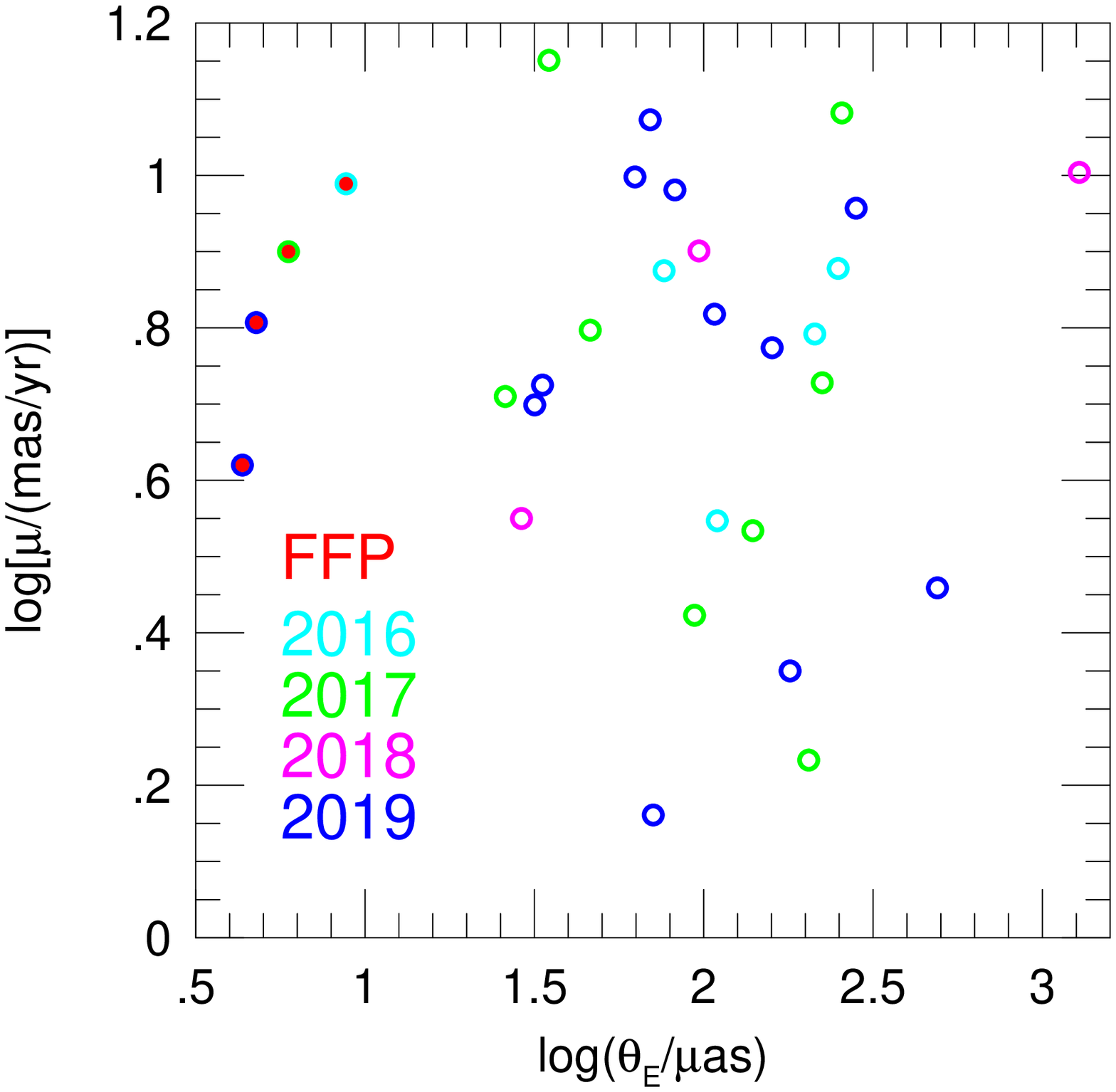}
\caption{Scatter plot of $\log\mu_\rel$ vs.\ $\log\theta_\e$ for the
30 FSPL giant-source events from four years of KMTNet data.  The events
are color coded by year: cyan (2016), green (2017), magenta (2018), 
and blue (2019).
The four FFPs at the left (with red interiors) all have $\mu_\rel > 4\,\masyr$,
whereas the 26 non-FFPs are distributed more broadly.  However no
strong conclusion can be drawn from this.
Note that Figure~9 of \citet{kb192073} corresponds
to the blue points in this Figure.
}
\label{fig:thstar}

\centering
\includegraphics[trim=0mm 10mm 20mm 10mm, width=90mm]{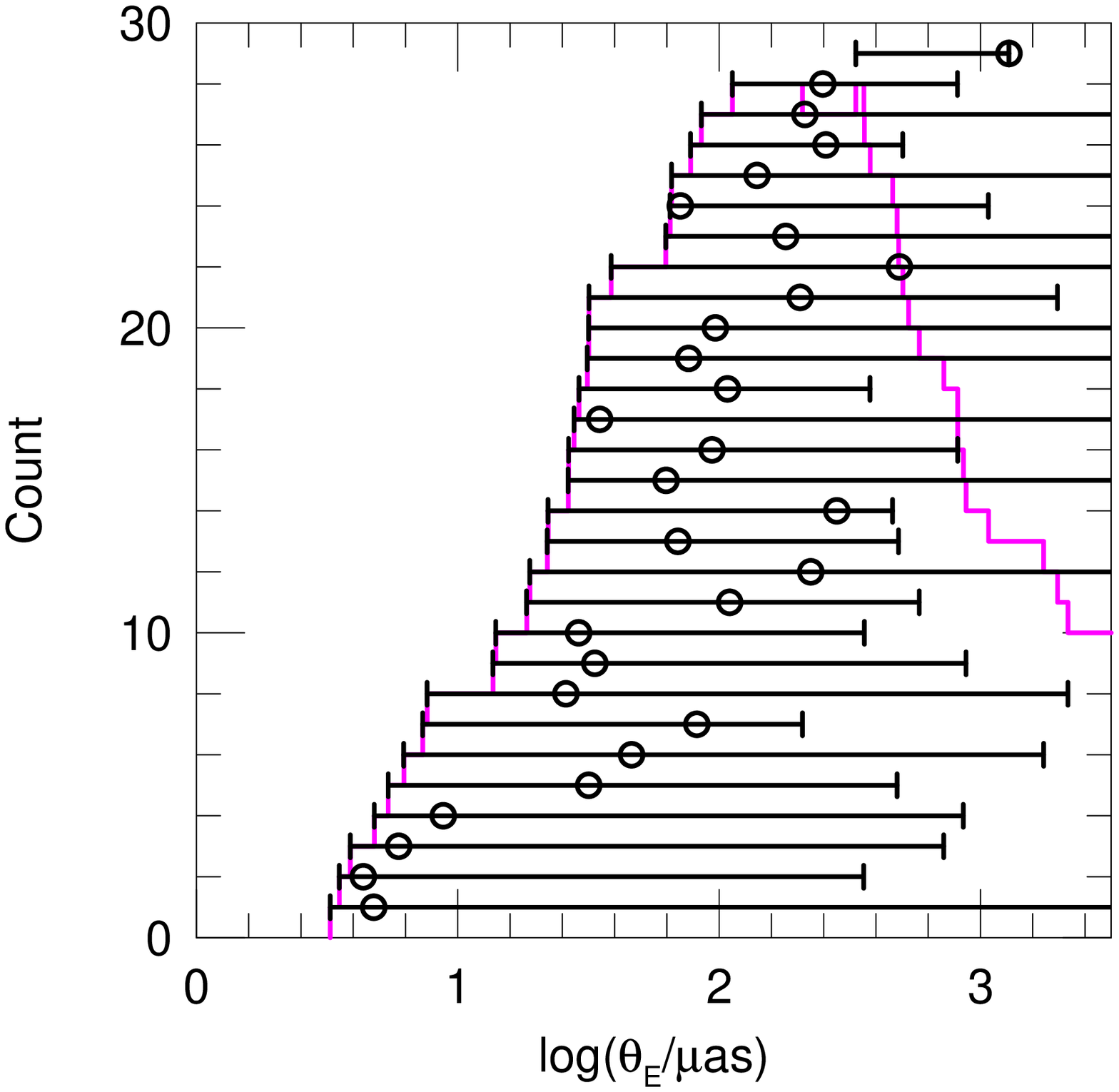}
\caption{Measurements (circles) and sensitivity limits (error bars)
of $\theta_\e$ for 29 FSPL events, rank ordered by lower limits.  As
the $\theta_\e$ of hypothetical events is progressively reduced below
that of the actual event, it eventually falls below the detection
threshold, either because $\Delta\chi^2_{\rm select}<1000$ (required to
detect the event) or $\Delta\chi^2_{\rm fspl}<20$ (required to be chosen
as an FSPL event).  As $\theta_\e$ is progressively increased, the
peak of the event eventually saturates, making it impossible to measure
$\theta_\e$.  KMT-2019-BLG-2528 with $\Delta\chi^2_{\rm select}=973$ is excluded
because the detection itself fails $\Delta\chi^2_{\rm select}>1000$.  The
magenta histogram shows the number of FSPL events that were sensitive to each
$\theta_\e$.
}
\label{fig:thetae_range}
\end{figure}

Figure~\ref{fig:thstar} shows a scatter plot of FSPL proper motions
against Einstein radii.  Generally, the proper motions are consistent
with those expected for bulge and disk lenses.  Note that the FFPs
are within this range, somewhat more tightly grouped but with a median
similar to the sample as a whole.

\section{{Selection Effects}
\label{sec:select}}

\subsection{Selection Effects Due to Lens Mass (thus, $\theta_\e$)}

There are two principal selection effects due to lens mass that can prevent a
given event from entering the sample. These derive from the
requirements that, first, the light curve must initially be selected as
a ``microlensing event'', and second, that
finite source effects must be detected in this event.
In Section \ref{sec:detect}, we adopted a threshold for finite source
effects: $\Delta\chi^2_{\rm fspl} \equiv \chi^2({\rm PSPL}) -
\chi^2({\rm FSPL})>20$.  The threshold for event detection,
$\Delta\chi^2_{\rm select}$, i.e., the difference in $\chi^2$ between
microlensing and non-microlensing interpretations of the light curve
as determined by EventFinder and the special giant-star searches,
varied somewhat for different subsamples
(see \citealt{kb192073}).  For purposes of this section, we impose
the minimum threshold that was common to all subsamples, namely,
$\Delta\chi^2_{\rm select}>1000$.  This has the effect of eliminating
one of the FSPL events from the sample, i.e., KMT-2019-BLG-2528 with
$\Delta\chi^2_{\rm select}=973$.  A hypothetical event
can drop out of the sample either because 
$\Delta\chi^2_{\rm fspl}$ falls below our threshold of 20 or because
$\Delta\chi^2_{\rm select}$ falls below our threshold of 1000.
Under our assumptions for real versus hypothetical events, 
we will see that both 
$\Delta\chi^2_{\rm fspl}$ 
and $\Delta\chi^2_{\rm select}$ will monotonically decline with falling mass.  

To understand how selection effects impact these two criteria, consider two
events that are ``identical'' in every respect except for the lens mass,
e.g., a real event and a hypothetical event with either larger or smaller
lens mass.  Here, ``identical'' means that the lens and source of the
hypothetical event both have the same 6 phase-space coordinates as the
real event and that the source radius and temperature are also the same.
Under these assumptions, $\mu_\rel$, $z_0=u_0/\rho$, $t_* = \rho t_\e$, and 
$t_\eff = u_0 t_\e$ are the same for the two events.  
However, because $\theta_\e$ scales as $M^{1/2}$,
 $t_\e$, $u_0$, and $\rho$ all differ.  Specifically,
$t_\e\propto M^{1/2}$,
$u_0\propto M^{-1/2}$, and
$\rho\propto M^{-1/2}$.  
We also assume that the observational sequences
of the real and hypothetical events are the same. Then, we can evaluate
upper and lower limits on $\theta_\e$ for each real event based on
comparison to hypothetical events under these assumptions.

First consider the case that the real and hypothetical events satisfy
$\rho_{\rm real} < \rho_{\rm hyp} \ll 1$, i.e., in particular, 
$M_{\rm hyp} < M_{\rm real}$.  Then, because $z_0$ is the same 
in both cases, both the PSPL and FSPL models
will have essentially the same morphologies in the peak region of each event.
In addition, because the duration of the transit,
$\Delta t_{\rm trans} = 2 t_\e\sqrt{\rho^2 - u_0^2}$, is invariant, they will 
have roughly the same number of data points over the peak region. On the 
other hand,  they will have a magnification ratio
$A_{\rm hyp}/A_{\rm real} \simeq (\rho_{\rm hyp}/\rho_{\rm real})^{-1} =
\theta_{\e,\rm hyp}/\theta_{\e,\rm real}$.  Hence, the real event will be 
brighter than the hypothetical one, and the number of
``signal photons'' (difference between the true number for the FSPL
model and the expected number for the incorrect PSPL model) will
fall linearly with $\theta_{\e,\rm hyp}$, while the number of background
photons remains the same.  Hence, $\Delta\chi^2_{\rm fspl}$ will fall
linearly with $\theta_{\e,\rm hyp}$ if the hypothetical event is ``above sky''
and quadratically if it is ``below sky''.  Once $\theta_\e$ falls sufficiently
that $\rho\ga 1$, the functional forms of the models in the two cases 
start to differ, but the principle remains the same.  We estimate 
$\Delta\chi^2_{\rm fspl}$ by scaling to the actual value found from the
FSPL and PSPL fits to the real data
and calculating the signal-to-noise ratio of each case
by assuming typical $2^{\prime\prime}$ FWHM seeing and a typical 
$I$-band background of 19.25 mag per square arcsec.

Next we turn to $\Delta\chi^2_{\rm select}$.
As long as $\rho\la 1$, the principal impact of declining $\theta_\e$ 
on $\Delta\chi^2_{\rm select}$ is that $t_\e$ is shorter, so that there are fewer
magnified data points for a given event. Hence, 
 $\Delta\chi^2_{\rm select}$ declines as 
$\Delta\chi^2_{\rm select}\propto t_\e\propto \theta_\e$.
Eventually, $\theta_\e$ becomes small enough that $\rho \gtrsim 1$, at which 
point, finite source effects dominate the morphology of the light curve. 
At this point, the effect of the further decline of 
$\theta_\e$ through $\rho=1$ 
on $\Delta\chi^2_{\rm select}$ is similar to that on $\Delta\chi^2_{\rm fspl}$.

On the other hand, as the mass is increased, eventually the peak flux 
$f_{\rm peak}\simeq 2f_s/\rho\propto M^{1/2}$ will grow so large that
the images become saturated over peak.  As discussed above, in many
cases it will still be possible to measure $\rho$ from the $V$-band images.
However, these are 10 times less frequent, and so for low-cadence fields
they may not sample the peak region.  Moreover, as the mass continues
to be raised, eventually the $V$-band images will also become saturated.
We model this process by assuming the same seeing and background as above,
and by reducing $\Delta\chi^2$ by a factor 20 for the regime in which
$I$ is saturated but $V$ is not.  Here, a factor 10 comes from the lower
cadence in $V$ band and 2 comes from the lower photon counts in the $V$ band.
This sets an upper limit on the detectable $\theta_\e$ for a given event.

In most cases, we do not actually know the lens mass, but we do know 
$\theta_*$ and $\theta_\e$.
Hence, we can still compare the real and hypothetical events in terms of
a measurable quantity, $\theta_\e = \theta_*/\rho$.  Then, for each event $i$,
with measured Einstein radius, $\theta_{\e,i}$, there is some range
$\theta_{\e,\min,i} < \theta_\e < \theta_{\e,\max,i}$ over which it could have
been detected, where $\theta_{\e,\min,i}< \theta_{\e,i}$ is set by the
$\Delta\chi^2$ threshold (either $\Delta\chi^2_{\rm fspl}$ or 
$\Delta\chi^2_{\rm select}$) and $\theta_{\e,\max,i}> \theta_{\e,i}$ is set
by saturation.  

Figure~\ref{fig:thetae_range} shows these ranges of sensitivity to $\theta_\e$
for the 29 FSPL events with $\Delta\chi^2_{\rm select}>1000$, rank ordered by
the $\theta_{\e,\min}$ of each event.  The most important feature of this
Figure is that the 4 lowest-$\theta_\e$ events (i.e., the 4 FFPs) are all
pressed up near their detection limit.  By contrast, and with one 
very telling exception, there is no such effect for the 
highest-$\theta_\e$ events.  For example, the 8 events with 
$\theta_\e> 200\,\muas$ occupy a broad range of positions relative to the
detection limits for that event.  This suggests that the fact that there
is only one event with $\theta_\e>500\,\muas$ is due to the low frequency
of such events in nature  rather than lack of sensitivity of the survey.
Indeed, as shown by the magenta curve, of the 29 FSPL events, 22
would be able to detect events with $\theta_\e=500\,\muas$.
Moreover, there is a well-understood reason for the paucity of 
$\theta_\e\ga 500\, \muas$ events: as discussed in Section~\ref{sec:expected},
these are almost entirely due to disk lenses, and in a mass range for which
the mass function is already in decline.
Indeed, the one event that lies above this limit 
(OGLE-2018-BLG-0705), is known to lie far in the foreground 
$D_L\sim 2.5\,\kpc$ by two independent arguments.  First, it has a measured 
$\pi_\e$  (in addition to the known $\theta_\e$), yielding 
$\pi_\rel = \theta_\e\pi_\e$.  Second, its {\it Gaia} proper motion indicates 
that it is a member of NGC 6544, whose distance is well-determined by several 
techniques.   See Appendix~\ref{sec:individual}.  Hence, the
fact that the low-$\theta_\e$ events are all pressed up against
the detection limit hints that these may represent the ``edge''
of a population that we are just barely detecting.

\subsection{Selection Effects Due to $\mu_\rel$}

There are also selection effects due to lens-source relative proper motion.
To understand these, we consider hypothetical events with the same
$\theta_\e$ and lens-source trajectory, but moving with faster or slower
proper motion.  For hypothetical proper motions that are higher,
the hypothetical event will be ``sped up''  
($t_\e$, $t_\eff$, and $t_\star$ will all be shorter) compared to the real one,
so $\Delta\chi^2_{\rm select}$ and $\Delta\chi^2_{\rm fspl}$ will both
fall inversely as $\mu_\rel^{-1}$, and so will eventually fall below
our thresholds of selection and/or detection.  On the other hand, as
$\mu_\rel$ is decreased, the event will be ``slowed down'', thus
reducing $\mu_{\rm thresh}$ in direct proportion, i.e., 
$\mu_{\rm thresh}\propto\mu_\rel$.
Thus, it will eventually fall below our search limit of
$\mu_{\rm thresh}>1\,\masyr$.  Our main concern here will be the proper-motion
selection effects for the 4 FFPs.  These have
$\theta_\e  = (4.35, 4.77, 5.94, 8.80)\,\muas$,
$\mu_\rel   = (4.17, 6.14, 7.95, 9.74)\,\masyr$,
$\mu_{\rm thresh} = (3.46, 11.17, 10.34, 8.80)\,\masyr$,
$\Delta\chi^2_{\rm select} = (2180, 2663, 2992, 6249)$, and
$\Delta\chi^2_{\rm fspl} = ( 577, 234, 187, 1513)$.  Therefore, they have,
respectively,
$\mu_{\rel,\min} = (1.21, 0.55, 0.77, 1.11)\,\masyr$ and
$\mu_{\rel,\max} = (9.09, 16.35, 23.79, 60.87)\,\masyr$.
The lower limits play very little role, but the upper limits will be 
important when evaluating the kinematic constraints on the FFP masses (see Section \ref{sec:constraints}).

\section{{Constraints on the FFP Population}
\label{sec:constraints}}

If the ``Einstein Desert'' is real, then it implies that the objects
that lie below this gap are a separate population of low-mass objects.
The existence of this gap will then be an important constraint on
the mass function of these objects.
Hence, the first question that must be addressed is,
how statistically secure is this gap?

\subsection{{Is the Einstein Desert Real?}
\label{sec:desert-real}}

The fact that the eye is drawn to the gap in Figure~\ref{fig:cum3} does 
not, in itself, make it real.  We can construct the following ``naive test''
by noting that there are 4 detections within a factor of 2 below the gap,
and 6 within a factor of 2 above it.  Given that the sensitivity of
the survey is monotonically increasing over this entire range and that
the gap itself covers a factor of 3, we would expect a continuous
distribution to generate $(4+6)/2\times \log(3)/\log(2)=7.9$ detections
in the region
of the gap.  The fact that none are detected has a formal probability of
$p= 0.0004$.  However, we call this test ``naive'' because it is constructed
after the fact.  The above statistical evaluation would be valid only if we
asked the question before conducting the experiment.  We might have
asked a dozen questions, such as: is there a gap around 
$\theta_\e\sim 15\,\muas$, around $30\,\muas$ etc.  Or perhaps the data
would have yielded some other peculiar feature, causing us to construct
some other posterior test designed to highlight its improbability.
If we could imagine constructing 50 such tests, then the probability of
finding $p=0.0004$ in one of them is 2\%.  This is still small but the
case would not be overwhelming.

Nevertheless, the fact is that we did not enter this investigation without
any prior knowledge.  \citet{mroz17} already found an analogous gap in
the Einstein timescale distribution at $t_\e\simeq 0.5\,$days.  At typical
lens-source relative proper motions $\mu_\rel\sim 7\,\masyr$ (for disk lenses), 
we would expect a gap at $\theta_\e = \mu_\rel t_\e \sim 10\,\muas$, which
is similar to the geometric center of the gap in Figure~\ref{fig:cum3}: 
$15\,\muas$.  Hence, the above statistical test should be taken approximately
at face value.  That is, we regard the suggestion by \citet{mroz17}
of a new population as confirmed\footnote{In fact, \citet{mroz17} adopted
exactly the cautious approach recapitulated above toward the evidence
that they presented for a new population.  The statistical significance
for this suggestion was roughly comparable to ours: they found 6 events below 
the gap (compared to our 4).  However,  with less complete light-curve coverage 
and no finite-source effects, these were individually less secure.  Their
(very appropriate) conservative orientation was reflected in their title
``No  large  population  of  unbound  or  wide-orbit  Jupiter-mass planets'',
which did not mention a new population.  The abstract introduces this
suggestion with the cautionary phrase ``may indicate''.}.

\subsection{{$\delta$-Function FFP Mass Function}
\label{sec:delta-function}}

When \citet{sumi11} and \citet{mroz17} first suggested their respective
FFP populations, they each presented their frequency 
estimates in terms of $\delta$-function mass functions.  Neither argued
that the FFP mass function was sharply peaked, but rather used
$\delta$-functions as a convenient way to characterize the basic mass scale
and frequency.  In Section~\ref{sec:power-law}, we will argue that,
especially in light of the Einstein radius measurements presented here,
power laws provide a more useful framework to characterize FFPs.
However, there is one constraint on the FFP mass function that is
most easily described in terms of $\delta$-functions, namely, that
the observed FFP proper motions put a lower limit on the FFP mass scale.

As noted in the discussion of Figure~\ref{fig:thstar}, the observed
distribution is consistent with that of typical disk lenses.  Strictly speaking,
however, this really applies only to disk lenses with $D_L\ga 2\,\kpc$
for which the component of the proper motion that is due to the
peculiar motion of the lenses (relative to Galactic rotation)
is small compared to the mean relative proper motion that is set
by the proper motions of the bulge sources.  For $D_L\la 1\,\kpc$,
by contrast, these peculiar motions dominate, and the expected amplitude
of the relative proper motion grows $\mu_\rel\propto D_L^{-1}$ with declining
distance.  Hence, at sufficiently low $M$ (hence, low $D_L$),
the observed ``typical disk'' proper motions of the FFPs should come
into strong tension with what is expected for nearby lenses.

We quantify this conflict as follows.  For each assumed lens mass $M$, and 
for each of the four FFP events, we calculate $\pi_\rel=\theta_\e^2/\kappa M$.
We adopt $\pi_S$ from \citet{nataf13} and so find $D_L=\au/(\pi_\rel+ \pi_S)$,
and we adopt $\bmu_{S,\hel}$ from {\it Gaia}.  We model the disk as having
a flat rotation curve with $v_\rot = 235\,\kms$.  We model the disk
lenses as having velocity dispersions $\sigma_l = 28\,\kms\sqrt{\eta}$ and 
$\sigma_b = 18\,\kms\sqrt{\eta}$, where $\eta = \exp(D_L/2.5\,\kpc)$,
and we assume an asymmetric drift of 
$v_\rot - \sqrt{v_\rot^2 -(18^2 + 28^2 + 33^2)\eta\,(\kms)^2}$.  We conduct
a 2-dimensional integral over this velocity distribution, first calculating 
$\bmu_{\rel,\hel}$ and than converting to $\bmu_\rel$.  At that point, we exclude
realizations that exceed the upper limits that are described in
Section~\ref{sec:select} for each event.  And, of course, we weight
the result by $\mu_\rel$.  We find the fraction of this predicted $\mu_\rel$
distribution that
have proper motions less than the observed values, $g_i$.  For a fair
sample, we expect that the $g_i$ should be uniformly distributed, with
mean $\langle g_i\rangle = 1/2$, but if the model is systematically
overestimating the proper motions (i.e., it assumes that the lenses are closer
than they actually are), 
then we expect $g_i\ll 1/2$.  Hence we adopt a likelihood
estimator $L=\prod_{i=1}^4 g_i$.  
We find that FFPs with $M<1.5\,M_\oplus$ are disfavored at $2\,\sigma$, while
those with $M<0.95\,M_\oplus$ are ruled out at $3\,\sigma$.  
These limits correspond
to mean lens distances of 1.1 kpc and 0.7 kpc, respectively.  That is,
these results conform to our naive expectation.

Thus, any compact mass function (not necessarily a $\delta$-function) for
which the expected detections were $M\la M_\oplus$ would be heavily
disfavored.  However, as we will see explicitly in Section~\ref{sec:power-law},
only a small fraction of expected detections from viable power-law distributions
are in this range.

\begin{figure}
\centering
\includegraphics[trim=0mm 10mm 20mm 40mm, width=90mm]{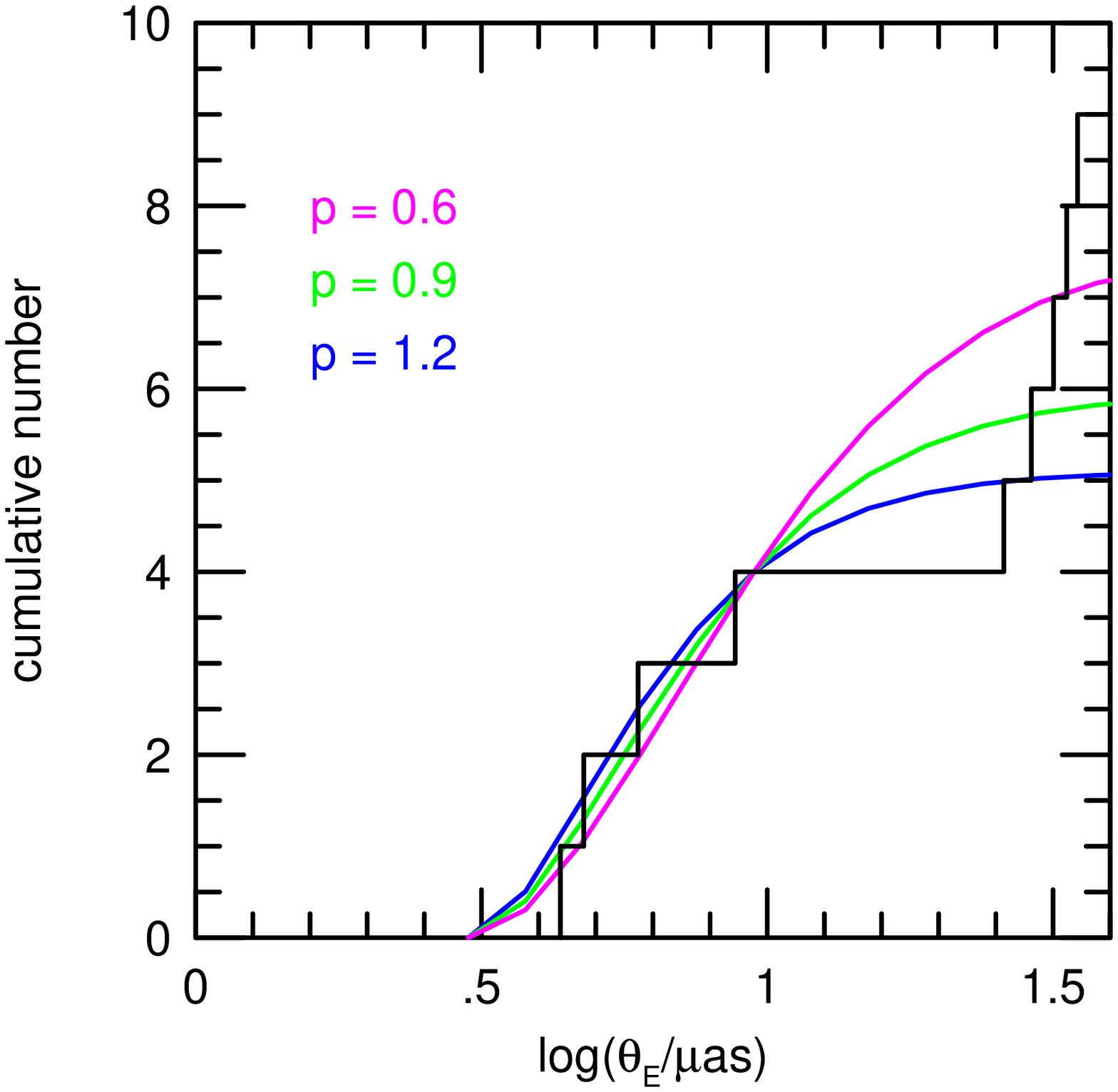}
\caption{Cumulative distributions of FFP detections for three power-law
indices (colored curves) compared to the actual detections (black).
The curves are each normalized to 4 detections at Einstein radius
$\theta_\e=9\,\muas$.  The three curves then predict (3,1.5,1) detections
within the observed Einstein Desert
for $p=(0.6,0.9,1.2)$, which argues strongly that the index is steeper
than the measured one for bound planets, $\gamma=0.6$.
}
\label{fig:comp}

\centering
\includegraphics[trim=0mm 10mm 20mm 20mm, width=90mm]{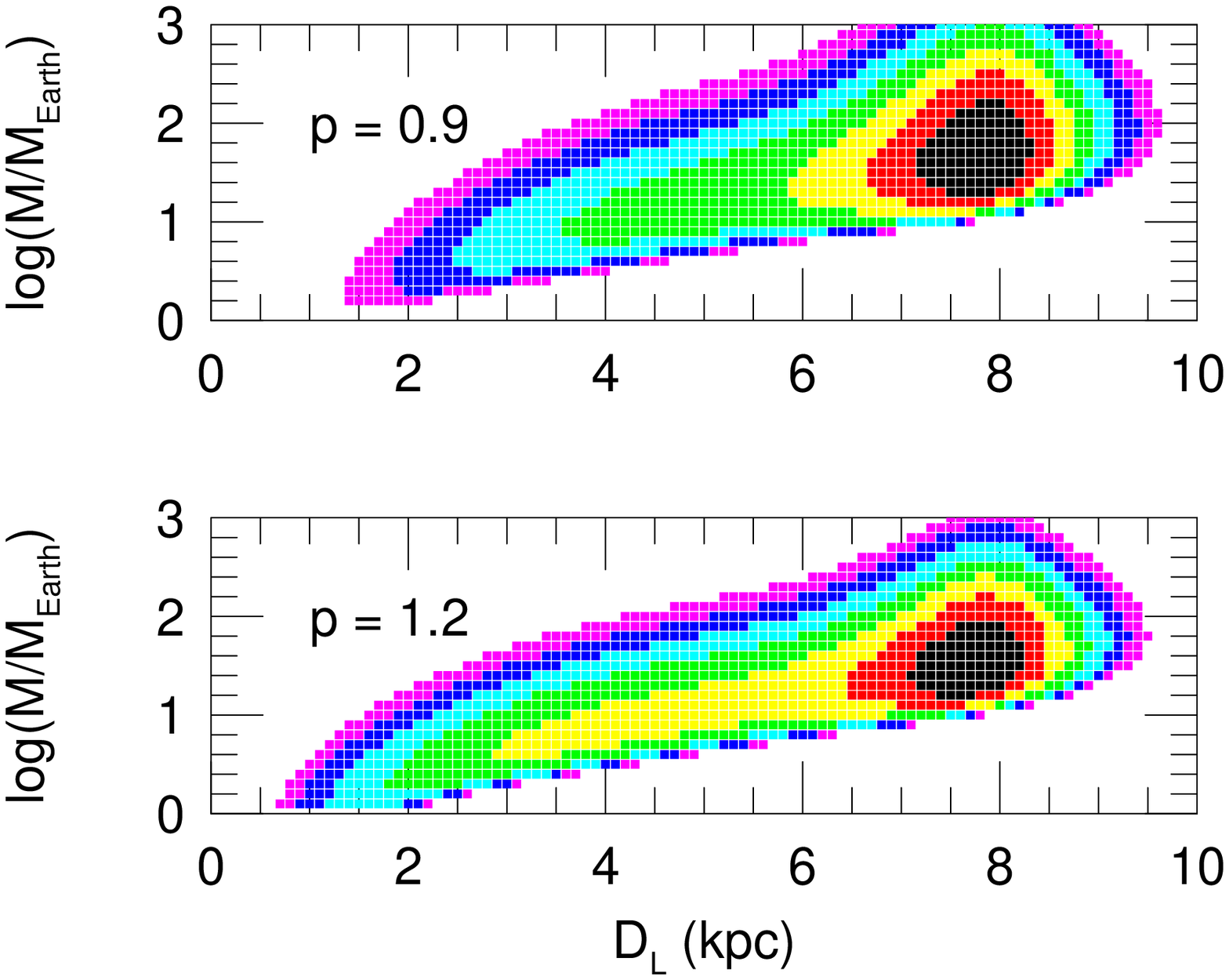}
\caption{Relative number of FFP detections predicted for $p=0.9$ and
$p=1.2$ power-law distributions, as a function of distance from the Sun
and FFP mass.  The contours represent steps of a factor 1.5.  For
$p=1.2$, the larger area in yellow and green almost perfectly compensates
the lower amplitude compared to the smaller black region, so that 
the FFPs are predicted to come equally from the bulge and the disk.
For $p=0.9$ the ratio is bulge:disk = 5:4.}
\label{fig:md}
\end{figure}

\begin{figure}
\centering
\includegraphics[trim=0mm 10mm 20mm 10mm, width=90mm]{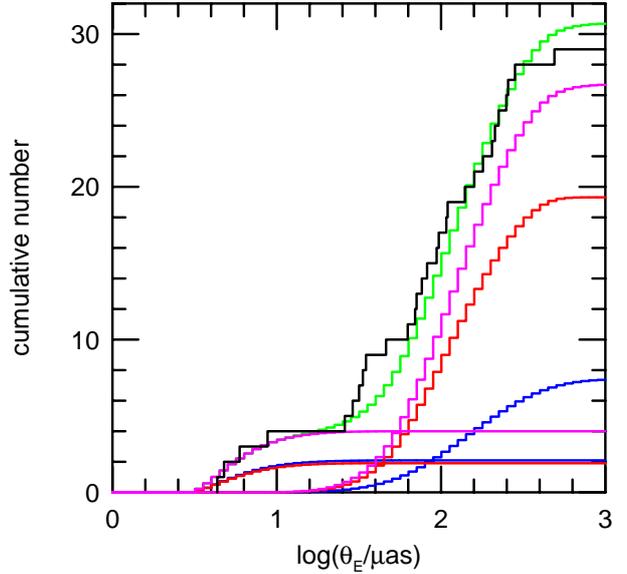}
\caption{Cumulative distribution of predicted FSPL detections (colored
curves) versus the actual detections (black)
of 29 events.  The FFP model ($p=1.2$), which flattens for 
$\log(\theta_\e/\muas)\ga 1.2$, is
normalized to 4 detections, shown as disk (blue), bulge (red), and combined 
(magenta) curves.  The same three colors are used for the predictions for
stars and BDs, which start to rise at $\log(\theta_\e/\muas)\sim 1.2$.
The star+BD normalization is set so that the overall prediction
(green) matches the observed profile.}
\label{fig:cumpred}
\end{figure}

\subsection{{Power-Law FFP Mass Function}
\label{sec:power-law}}

The FFP mass function is likely to be better characterized by
a power-law function,
\begin{equation}
{dN\over d\log M} = Z\times \biggl({M\over M_{\rm norm}}\biggr)^{-p},
\label{eqn:power-law}
\end{equation}
than a $\delta$-function. Like the latter, a power-law function
requires two parameters, namely the power-law index $p$ and the
normalization $Z$, while the reference mass $M_{\rm norm}$ is an
arbitrary zero point whose inclusion allows $Z$ to have simple units;
i.e., (dex)$^{-1}$.  The first argument favoring a power-law distribution
is that there is no reason to expect that FFPs
will have special mass scale, only that selection biases may possibly
``pick out'' FFPs of a certain scale.  Second, FFPs are very likely
to be related in some way to bound planets, whose mass-ratio function
is very well characterized by a power law.  For lenses with an observed
$\theta_\e$, the inferred mass varies as a function of distance,
$M =\theta_\e^2/\kappa\pi_\rel$, with $\pi_\rel = \au(D_L^{-1}-D_S^{-1})$,
and so can be very different if the lenses are assumed to be in the
disk or bulge.  Only a continuous mass distribution (of which, a power-law
function has the simplest form) can handle these possibilities simultaneously.

Moreover, we do in fact have some prior information on the power-law index $p$.
For planets in the mass range under consideration,
$M< M_{\rm Saturn}$, it is difficult to imagine any formation scenario
other than within the disks around stars, i.e., the same process
that gives rise to the known population of bound planets found by
microlensing planet searches.  These have a power-law mass-ratio
distribution, $dN/d\log q\propto q^{-\gamma}$, with $\gamma\simeq 0.6$
(Jung et al., in preparation, Zang et al., in preparation, 
\citealt{suzuki16}, \citealt{shvartzvald16}).
The FFPs must either be drawn from this population and ejected to larger
and/or unbound orbits, or they must form in the outer portions of
their solar system, i.e., beyond the range of detection of the known
bound population.  In either case, we would expect this to result
in a mass distribution with $p>\gamma$.  That is, the probability
for a planet drawn from the known bound population to be scattered
to a large or unbound orbit monotonically increases with declining mass
because this generally requires scattering off a heavier planet.
And planets formed in the outer disk (by core-accretion rather than
gravitational collapse) should be heavily biased toward low mass
because of the paucity of raw material and the long dynamical timescales.
For example, in our own Solar System only dwarf planets such as Pluto
are believed to have formed in situ in the outer Solar System, while
Uranus and Neptune are believed to have formed inside the current orbit of
Saturn and to have been scattered to their present-day orbits.

In the observed mass-ratio function for bound planets, the number of
planets with mass ratio greater than a certain $q$, is $\propto q^{-\gamma}$.
Then, assuming that scattering to large or unbound orbits is proportional
to the number of such potential scatterers, $p=k\gamma$, with $k=2$.
In practice, heavier planets will be more efficient scatterers, which
would imply $1<k<2$.  On the other hand, if the number of FFPs is of order
or larger than the number of bound planets (as seems to be the case,
see Section~\ref{sec:bound}), then the original value of $\gamma$
must have been larger than currently observed, with lower mass members
of the original bound population having been preferentially removed.
In this case, the present-day observed mass ratio
distribution (and, in particular, its power-law form) must reflect the
imprint of the scattering process in addition to the formation process.
However, because the FFP population is not overwhelmingly dominant,
it is unlikely that taking account of formation-plus-scattering can
drive $k$ much above the range $1<k<2$.

Starting from these very general considerations, we ask 
whether and how the FSPL FFP sample can further constrain $p$.  
Figure~\ref{fig:comp} shows the predicted cumulative distribution of
FFPs based on the range of power laws described above, $p=(0.6,0.9,1.2)$,
corresponding to $k=(1,1.5,2)$.  In making these predictions we have
adopted relative detection efficiencies 
${\cal E} = (4/3)X$ $(X<0.5)$, 
${\cal E} = (2X +1)/3$ $(0.5<X<1)$,
${\cal E} = 1$ $(1<X<2)$,
${\cal E} = 2-X$ $(2<X<3)$,
with $X\equiv\log(\theta_\e/3\,\muas)$.
In the regime $X>0.5$, in which the detections are roughly randomly
distributed in the range of detectability (Figure~\ref{fig:thetae_range}),
this function approximates the relative fraction of events that
are sensitive to a given $\theta_\e$.  For $X<0.5$, we complete this
function linearly by imposing a threshold at $\theta_\e=3\,\muas$, which
is supported by the fact that all four FFPs are pressed up close to this limit.

The curves are all normalized to the 4 detections with $\theta_\e<9\,\muas$.
As can be seen, all three curves account of the form of the detections
very well.  However, they then predict, respectively, about 3, 1.5, and 1
event in the Einstein Desert.  Hence, they are, respectively,
strongly disfavored, mildly disfavored, and acceptable.  

In Figure~\ref{fig:md} we show the predicted distribution of detections
as a function of FFP mass and distance for the latter two models.
The contours change color at factors of 1.5.  The bulge FFPs are centered at 
$M\sim (40,30)\,M_\oplus$ for $p=(0.9,1.2)$, while the disk FFPs are
at a broad range of lower masses, as expected from the facts that $\theta_\e$
is measured for the objects in our sample 
and $M=\theta_\e^2/\kappa\pi_\rel$.  As we will soon see,
there are almost exactly the same number of bulge and disk FFPs in the
$p=1.2$ model: the fact that the yellow and green contours are much
larger than the black contour compensates for their 2.2--3.4 times 
lower amplitude.  For the $p=0.9$ model, we find that the ratio
is bulge:disk = 5:4.

To normalize this distribution we predict the stellar and brown-dwarf
FSPL events using a Chabrier initial mass function (with bulge stars 
that have $M>1\,M_\odot$ being converted into remnants).  
See Figure~\ref{fig:cumpred}.  
We normalize the FFP curves to 4 total
detections and normalize the stellar curve to approximately match the
observed distribution.  The bulge and disk FSPL events are shown
in red and blue respectively, with magenta showing the sum of these.
There are two sets of such curves, one for FFPs and the other for
stars and BDs.  The green curve shows the overall sum.  For the FFPs,
the bulge and disk curves are almost identical.  As noted above, for
the $p=0.9$ model, the bulge FFPs would be favored 5:4.  From the
relative normalization of the FFP and star-plus-BD curves,  the FFP
mass function can be expressed as
\begin{equation}
{dN\over d\log M} = {0.39\over{\rm dex\times star}}\times 
\biggl({M\over 38\,M_\oplus}\biggr)^{-p},\
\quad (p=0.9\ {\rm or}\ 1.2).
\label{eqn:power-law2}
\end{equation}
Because the mean mass of the adopted stellar-BD mass function is 
$\langle M\rangle = 0.199\,M_\odot$, this can also be written
\begin{equation}
{dN\over d\log M} = {1.96\over{\rm dex}\times M_{\odot,\rm stars}}\times 
\biggl({M\over 38\,M_\oplus}\biggr)^{-p}.
\label{eqn:power-law3}
\end{equation}
The uncertainty in the normalization is strongly dominated by the Poisson error
of 4 FFP detections, i.e., it is 50\%.  We consider that it is unlikely
that the power-law index lies far outside the indicated range 
$0.9\la p \la 1.2$.  For indices much below $p=0.9$, the FFPs would
populate the Einstein Desert.  While there are no strong constraints
on power laws that are substantially steeper than $p=1.2$, it is hard
to imagine a mechanism that would generate them.

Assuming that these power laws apply to FFPs from Earth to Saturn
masses, then this range contains a total of 
$130\,M_\oplus$ or $200\,M_\oplus$ of FFPs per solar mass of stars
for the $p=0.9$ and $p=1.2$ models, respectively.

We note that the form of the bulge and disk stellar FFP
detections shown in Figure~\ref{fig:cumpred}
are in reasonable accord with the expectations of
Section~\ref{sec:expected}.  The observed FSPL events are above
the expectations in the BD regime.  This may reflect an underestimation
of the BD component in the Chabrier mass function, but it could also
reflect a statistical fluctuation.  Addressing this question would
take us beyond the scope of the present investigation.

\section{{Discussion: Relation of FFPs to Known Populations}
\label{sec:discuss}}

When we introduced the acronym ``FFP'', we were careful to have it refer to
``candidates'', rather than just ``free-floating planets''.
The main reason for this is the general concern that the low-mass objects that
appear ``isolated'' in microlensing events could be bound planets in orbits
that are too wide to enable microlensing signatures of the host.
This possibility has always been recognized and was
explored in depth by \citet{clanton14,clanton14b,clanton16}.
\citet{gould16} pointed out that the orbital separations of
FFPs in very wide (``Kuiper-like'') and extremely wide (``Oort-like'') orbits
could eventually be measured.

Here, we compare the FFP population as derived from our FSPL survey,
as characterized by Equations~(\ref{eqn:power-law2}) and (\ref{eqn:power-law3})
to various known populations, in order to obtain a more comprehensive
picture.  This will include planets discovered by microlensing in typical
orbits, known planets in very wide orbits, and known interstellar objects.

We adopt the orientation that the most likely origin of FFPs is that
they they formed within a factor of a few of the snow line, a region
that was rich in proto-planetary materials, and where dynamical timescales
were relatively short, and that they were ejected by dynamical processes,
either to much wider orbits or to unbound orbits.  Hence, we regard it as likely
that FFPs comprise both bound and unbound objects.  Within this framework,
we expect a greater fraction of high-mass FFPs to be in bound orbits,
simply because it is easier to eject a test particle than an object whose
mass is a substantial fraction of that of the perturber.  

\subsection{{Comparison to Mroz et al.\ (2017)}
\label{sec:mroz2017}}

\citet{mroz17} considered a $\delta$-function model with FFP mass
$M=5\,M_\oplus$  to account for their
$N_{t_\e}=6$ short-$t_\e$ events.  They 
obtained a best fit of $R_{5\,M_\oplus,t_\e}=10$ FFPs per star.  
Although they did not quote an error in this estimate, the
Poisson noise yields 
$\sigma_{5\,M_\oplus,t_\e} = R_{5\,M_\oplus,t_\e}/N_{t_\e}^{1/2}=4.1$.

To compare our result to theirs, we insert the same $\delta$-function model 
into the formalism described in Section~\ref{sec:power-law},
thereby obtaining $R_{5\,M_\oplus,\theta_\e}=15.3\pm 7.6$, based on our
$N_{\theta_\e}=4$ small-$\theta_\e$ detections.  Hence, the two determinations
are consistent at the $0.6\,\sigma$ level.

Although we mainly restrict consideration to our own results in this
work, we note that, because the two normalizations are consistent, 
it can also be appropriate to combine them.  We then find that the
combined normalization is a factor
\begin{equation}
F_{\rm renorm} = 0.73\pm 0.24
\label{eqn:renorm}
\end{equation}
smaller compared to results from our study alone.  Note that the fractional
error is then $\simeq 1/3$ rather than 1/2, as would be expected from the
fact that there are a total of 10 detections, rather than 4.
Note also that this renormalization would apply to both 
Equations~(\ref{eqn:power-law2}) and (\ref{eqn:power-law3}) 
and to either indicated power-law index.

\subsection{{Comparison to the Known Bound-planet Mass-ratio Function}
\label{sec:bound}}

At present, the masses of most microlensing planets are unknown.
Hence, what is most precisely measured is the planet-host mass-ratio ($q$)
function, which can be expressed, in analogy to Equation~(\ref{eqn:power-law}),
\begin{equation}
{dN\over d\log q} = Z_q\times \biggl({q\over q_{\rm norm}}\biggr)^{-\gamma},
\label{eqn:mass-ratio}
\end{equation}
which, using a zero point $q_{\rm norm}=10^{-3.5}$, has measured values
$Z_q = 0.175\,({\rm dex})^{-2}$/star and $\gamma=0.6$
(Jung et al. 2022, in preparation, Zang et al. 2022, in preparation).
There are three things that make the comparison difficult.  First,
the units of $Z$ are $({\rm dex})^{-1}/$star while 
the units of $Z_q$ are $({\rm dex})^{-2}/$star.  The additional ``dex'' means
``per decade of projected separation''.  We account for this by
regarding two decades of separation (roughly 0.1--10 $\au$)
as being approximately representative of the whole population. 
That is, we roughly account for not only the cold planets detected by
microlensing but also the hot and warm planets detected by transit and
radial velocity surveys.  Second, the denominator term ``star'' does not
mean exactly the same thing for the two power-law functions.  For our FSPL
study, events enter the sample in proportion to their number density,
but for the mass-ratio functions studies of Jung et al. (in preparation) and
Zang et al. (in preparation), they enter as microlensing
events, whose frequency is weighted by their cross section, i.e., 
$\propto M^{1/2}$.  Therefore, if we want to convert 
Equation~(\ref{eqn:mass-ratio}) to ``per solar mass of stars'', we should
calculate the mean mass weighted by $M^{1/2}$, i.e.,
$\langle M\rangle_{\sqrt{M}}\simeq 0.33\,M_\odot$.  Third, eponymously,
the argument of the mass function is mass, while the argument of the
mass-ratio function is mass ratio.  We account for this by evaluating
at bound-planet mass $m_p=q\langle M\rangle_{\sqrt{M}}$.  With these 
approximations, we infer a frequency of known bound planets of
\begin{equation}
{dN\over d\log m_p} = {1\over{\rm dex}\times M_{\odot,{\rm stars}}}\times 
\biggl({m_p\over 38\,M_\oplus}\biggr)^{-0.6}.
\label{eqn:bound-mass}
\end{equation}

Comparison of Equations~(\ref{eqn:power-law3}) and (\ref{eqn:bound-mass})
indicates that at the normalization mass there are 2 times more FFPs
than bound planets, while at lower masses, the ratio is even larger.
For example, at $m_p = 10\,M_\oplus$, it is about 3 for the $p=0.9$ power
law.  Of course, this comparison required several approximations
due to the intrinsic incommensurablity of the underlying measurements.
Nevertheless, one can infer robustly that the number of FFPs is of
order or larger than the number of bound planets in the mass range
to which the FFP measurement is directly sensitive,
$5\la M/M_\oplus \la 60$ (see Figure~\ref{fig:md}).  This implies,
as foreshadowed in Section~\ref{sec:power-law}, that the form of
the bound-planet mass function was shaped partly, or perhaps mainly,
by the ejection process, rather than just the formation process.

\subsection{{Comparison to Known Uranus- and Neptune-like Planets}
\label{sec:doppel}}

In the entire Universe, there are exactly three known planets that
can be securely described as ``Uranus-like'' ( or ``Neptune-like''),
in that their masses and separations are of the same order as these two planets,
namely, Uranus and Neptune themselves, as well as OGLE-2008-BLG-092Lb, 
which has $s=5.3$ and $q=2.4\times 10^{-4}$ \citep{ob08092}.

\subsubsection{{Uranus and Neptune}
\label{sec:uranus}}

We focus first on Uranus and Neptune.  Equation~(\ref{eqn:power-law3})
predicts that, for $p=0.9$ and within a factor $\sqrt{10}$ of
$M=(M_{\rm Uranus}+M_{\rm Neptune})/2 = 16\,M_\oplus$, each solar mass of stars
should be accompanied by an average 4.3 FFPs (whether unbound or in wide 
orbits).  This can be compared to the two such observed planets in the Solar 
System, the only planetary system for which we have a complete census.  These
numbers are compatible within the Poisson error, and they would be more
so if we applied the renormalization given by Equation~(\ref{eqn:renorm}),
by which the expectation would be reduced to $3.1\pm 1.0$.

Furthermore, if a doppelganger of the Solar System
gave rise to a microlensing event at $D_L\sim 5\,\kpc$ in which either
its ``Uranus'' or its ``Neptune'' generated a characteristic FSPL profile,
there is only a small chance that its ``Sun'' would leave a trace on
the event.  That is, if a planet lies at (3-dimensional) physical separation 
from its host, $r$, then the probability that its normalized projected
separation, $s$, and host-source impact parameter, $u_0$, lie within
specified regimes are,
\begin{equation}
p(s<s_{\max}) = 1-\sqrt{1-{s^2\over s_{\rm full}^2}},\qquad
p(u_0<u_{0,\max}) = {u_0\over s_{\rm full}},
\label{eqn:range}
\end{equation}
where $s_{\rm full}\equiv r/D_L\theta_\e$.  We note that for $D_L=5\,\kpc$
and $D_S=8\,\kpc$, $D_L\theta_\e= 3.91\,\au(M/M_\odot)^{1/2}$, but for the
entire range, $2.5\,\kpc<D_L<5.5\,\kpc$, this formula remains accurate
to $<4\%$.  Thus, for Uranus and Neptune doppelgangers, 
$s_{\rm full}=4.9$ and 7.7, respectively.  We focus first on Neptune.
According to Equation~(\ref{eqn:range}) there is a small ($p=8\%$) chance
that $s<3$, in which case, even if the host were detected, the planet
would be regarded as having a Saturn-like (not Neptune-like) orbit.  For FSPL
events, the main way that the host would be detected is that the source would
come within $u_0<u_{0,\max}\sim 2.5$, which would induce a peak magnification
$A\ga 1.03$.  Overall, this occurs with probability $p=32\%$, but
about 7\% is due to the cases with $s<3$, leaving $p=25\%$ that could
in principle be recognized as having Neptune-like orbits.  At the median
projected separation of the $s>3$ population, namely $s_{\rm med} = 6.8$,
the encounter with the host would occur 
$\Delta t \simeq (s_{\rm med}/s_{\rm full})r/D_L\mu_\rel \rightarrow 
68\,{\rm day}(\mu_\rel/6\,\masyr)^{-1}$ before or after $t_0$.  To reliably 
identify a peak in such a low-amplitude event, we should require that the
observations continue a factor 1.3 beyond peak, i.e., to 77 days.
Roughly 1/3 of all events will fail this criterion because the
peak region lies wholly or partly outside the microlensing season.
Thus, we expect about 16\% of true bound Neptunes that are detected as FSPL
FFP events to be recognizable as such.  Repeating the same calculation
for Uranus yields 32\%.
Thus, the average probability for Uranus and Neptune is about 25\%.
Hence, with four examples, we should expect one host detection
assuming that all were bound.  However, the probability for zero
detections under this assumption is $p = 32\%$.  Moreover, there is
added uncertainty in this estimate due to the fact that the majority
of potential hosts are much less massive than the Sun.  If typical
separations of Neptune-like planets scale $\propto M^{1/2}$, then the above
calculations hold exactly.  If they they scale $\propto M$, then the
probability of host detections goes up, while if these typical separations
are independent of host mass, then the probability goes down.

The main implication of this exercise is that, at least regarding the
region of parameter space in which Equation~(\ref{eqn:power-law3}) is
directly sensitive to the observations, the Solar System is 
consistent with the hypothesis that all FFPs are due to bound planets
in wide orbits, such as those of Uranus and Neptune.

\subsubsection{{OGLE-2008-BLG-092Lb}
\label{sec:ob08092}}

Among bound planets with secure $s$ and $q$ measurements: 
OGLE-2008-BLG-092Lb is unique:  there are no other planets with
$s>3$ and $q<10^{-1.8}$.  Planets with $s<3$ are unlikely to be
in wide orbits, while those with $q>10^{-1.8}$ could in principle
have formed through another channel, i.e., gravitational collapse.

However, in addition to this one secure Neptune-like planet,
there is another bound planet, OGLE-2011-BLG-0173Lb, with
best estimates $s=4.7$ and $q=4.5\times 10^{-4}$ \citep{ob110173}.
In fact, there is an alternate solution with $s=0.22$ with virtually
identical $\chi^2$.  Nevertheless, based on the general statistical
properties of microlensing planets, \citet{ob110173} argued that
the wide solution was preferred.

\citet{wide-orbit} conducted a systematic search for wide-orbit
planets in 20 years of OGLE data, which recovered these two planets
(and no others in the ``Neptune-like'' domain).  They re-evaluated
and confirmed the argument that the wide solution for
OGLE-2011-BLG-0173Lb is preferred, based on updated microlensing statistics.
Including both planets, they derived a rate of $1.4^{+0.9}_{-0.6}$
``wide-orbit ice giants'' ($2<s<6$, $-4<\log q < -1.5$),
based on a total of 5 detections, only the above
two of which are in the regime of interest here.

To obtain a proper comparison to the FSPL sample, we 
apply their best-fit parameters $(A,n,m)=(1.04,1.15,1.09)$ to
their Equations (6) and (8), but restricted to 
($4<s<6$, $-4<\log q < -3$), and find an expected number (per star)
of $N=0.67\pm 0.47$, where we have adopted a Poisson error 
based on two detections.  To maintain consistency with our
other comparisons, we should convert this to expected Neptune-like
planets per solar mass of stars, i.e., $2.0\pm 1.4$.  The best-fit
value is identical to the actual number of such planets in the Solar System,
and, additionally,
there is a substantial Poisson error in this prediction.  Hence, this result
is likewise consistent with all Neptune-like FFPs being due to wide-orbit
bound planets.

\subsubsection{{Caveats}
\label{sec:caveats}}

Although both of the above tests indicate consistency with the
hypothesis that all of the FFPs in our FSPL sample are due to known
populations of wide-orbit bound planets, there is no strong evidence that
even a majority (let alone all) are due to such planets.  First,
the solar system is known to be rich in gas giant planets in ``Jupiter-like''
or ``Saturn-like'' orbits relative to the field.  From the mass-ratio
function, we would expect only 0.2 such planets within $\sqrt{10}$ of
their mean mass, compared to two observed.  Moreover, the presence of
Jupiter and Saturn is significant not only in that they may indicate
that the Solar System is planet-rich, but also in that they were implicated
in the expulsion of Uranus and Neptune to wide orbits.  Finally, the
observed characteristics of the Solar System may be subject to
subtle selection effects due to the existence of the observers.
For example, the turmoil in the Solar System engendered by the expulsion
process may have played a crucial role in the delivery to Earth of water
and/or organic materials, or perhaps to other conditions on Earth
that were crucial to the development of intelligent life.

Similarly, the estimates of Neptune-like planets from the detection of
wide-orbit planets are subject to several caveats.  First, the Poisson
error in this estimate is already quite large.  Second, one cannot
really be certain that the wide-orbit solution for OGLE-2011-BLG-0173Lb
is correct.  This solution is favored statistically only because
there are more planets with $s>1$ than $s<1$, but the power law
relation is derived from a sample that does not contain any planets
(other than OGLE-2008-BLG-092)
that are close to its observed value: $|\log s|\simeq 0.7$ and that
are also $\log q<-1.8$ (so not subject to alternative formation
mechanisms).  If this planet is eliminated, then the mean expected
number drops by a factor 2 and the fractional Poisson error becomes
much larger.

What we can say robustly is that there is a known population of bound
planets that contributes significantly to FSPL events that have been
(or possibly will be) in the FFP domain.  However, whether this is 10\%
or 100\%, we cannot now say.

Finally, we note that (contrary to the case of the Jovian FFPs
suggested by \citealt{sumi11}) there are no limits on bound Neptune-like
planets from direct imaging because they are too faint.

\subsection{{Comparison to 'Oumuamua-like Objects}
\label{sec:oum}}

'Oumuamua \citep{oumuamua} was discovered via the PanStarrs survey
as an asteroid-like object
that was leaving the Solar System in a strongly unbound orbit
(eccentricity $e=1.2$).  It may
or may not be related to FFPs, depending on its physical origin.
The conventional view, and the one that we will adopt here, is that
it is an interstellar asteroid that was formed in another solar system.
However, other ideas for its origin have been suggested, including
that it is an alien spacecraft \citep{loeb-oum} and that it is the
result of a violent collision of Solar System objects that occurred
inside the orbit of Mercury (B.\ Katz et al, 2018).

\citet{do-oum} calculated the space density of objects with
exactly the same physical properties as 'Oumuamua under the assumption
that the expected rate of detection was exactly 1 per 3.5 years
of PanStarrs operations as $n\simeq 0.20\,\au^{-3} =1.75\times 10^{-15}\,\pc^{-3}$.
To account for the fact that there have been no further reports of such objects
during the past 4 years,
we reduce this to $n = 1\times 10^{-15}\,\pc^{-3}$.  Following \citet{rafikov-oum}
we adopt physical dimensions (180m$\times$18m$\times$18m), and a typical
asteroid density of 3 gm/cm$^3$ to derive a mass 
$M_{\rm Oum} = 1.7\times 10^{11}\,{\rm gm}= 3\times 10^{-17}\,M_\oplus$, and
so a spatial density $\rho_{\rm Oum}=0.03\,M_\oplus\,\pc^{-3}$.  In order to
compare this to our FFP estimates, we first adopt a local stellar mass
density of $\rho_{\rm stars} = 0.03\,M_\odot\pc^{-3}$ and assume that this
detectability applies to 1 dex of object mass, to obtain
\begin{equation}
{dN_{\rm Oum}\over d\log M} = {3.3\times 10^{16}\over{\rm dex}\times
M_{\odot,{\rm stars}}},
\label{eqn:oum}
\end{equation}
implying an asteroid mass density per decade of 
$1.0\,M_\oplus/{\rm dex}/M_{\odot,{\rm stars}}$.

The main point to notice about this calculation is that this mass fraction
per decade (measured at $M=3\times 10^{-17}\,M_\odot$) is remarkably similar
to the value implied by Equation~(\ref{eqn:power-law3}) at its pivot point
($M=38\,M_\oplus$) of $32\,M_\oplus/{\rm dex}/M_{\odot,{\rm stars}}$.
That is, the two are consistent with being part of the same distribution
provided that the distribution function is approximately flat in mass, i.e.,
$p\sim 1$.  As we have already concluded that $p\sim 1$ based on
FFP detections, combined with the lack of detections in the Einstein Desert,
this suggests that FFPs and 'Oumuamua-like objects may result from the
same physical processes, i.e., the processes of planet formation and early
dynamical evolution.

In fact, \citet{do-oum} considered and rejected the idea that
'Oumuamua-like objects were created and ejected as part of planet formation,
primarily because they appeared to be too abundant.  However, we consider
that the error in the above estimate is an order of magnitude (i.e., 1 dex).
In particular, if 'Oumuamua-like objects were a factor 10 less common than
the \citet{do-oum} estimate, there would still be a 10\% chance that one would
be detected, which hardly can be considered to be an implausible scenario.

If we assume that they are part of the same process and that they participate
in the same power-law distribution, then by comparing the two measurements
separated by 18.1 dex in mass, which differ in amplitude by $-1.5\pm 1$ dex,
we can estimate a power-law index of
\begin{equation}
p = 1 + {-1.5\over 18.1}\pm {1\over 18.1} = 0.92\pm 0.06 .
\label{eqn:oum-power}
\end{equation}

However, it is also possible that the similarities of the mass densities
of FFPs and 'Oumuamua-like objects is just a coincidence and that
'Oumuamua arose from some entirely different process.  In addition
to the possibilities mentioned above, \citet{rafikov-oum} has suggested that
'Oumuamua results from the tidal disruption of a rocky planet by its
white-dwarf host.  There could be other possibilities as well.

Nevertheless, the hypothesis of a common origin can be subjected
to further tests.  If more 'Oumuamua-like objects are found by 
improved surveys, their frequency and power-law distribution can
be better measured.  If this power law is consistent with the $p\sim 1$
that is implied by the common-origin hypothesis, this would constitute
further evidence.  Similarly, better measurements of the FFP power
law could confirm (or contradict) its apparent agreement with the
one needed to connect the FFP and 'Oumuamua measured frequencies.

In this regard, it is important to note that there has already
been a detection of a second extra-solar 
minor body, 2I/Borisov \citep{higuchi20},
of substantially different mass, $M\sim 5\times 10^{-14}\,M_\oplus$
\citep{jewitt20}, i.e., 1700 times more massive than 'Oumuamua.
Unfortunately, to our knowledge, there is not as yet a 
published estimate of the space density of objects of this class.

\subsection{{Summary}
\label{sec:summary}}

The comparison to various known populations can be summarized as follows.
First, the frequency of FFPs derived from our FSPL sample is consistent
with that from the PSPL sample of \citet{mroz17} at the $1\,\sigma$ level.
Combining the two leads to pre-factors of $0.28\pm 0.09$
and $1.43\pm 0.47$ in
Equations~(\ref{eqn:power-law2}) and (\ref{eqn:power-law3}), respectively.
Second, the frequency of FFPs is of order or larger than that of
known bound planets.  Third,
if the Solar System's endowment of planets with similar masses and
orbits to those of Uranus and Neptune, then these objects are likely to
contribute a substantial fraction of our detections.  However, there
is some evidence that the Solar System may not be typical.  Fourth, the
observed FFPs and 'Oumuamua-like objects are consistent with being
drawn from the same power-law distribution, in which case its index
would be $p=0.92\pm 0.06$.

\appendix
\section{{Remarks on Individual Events}
\label{sec:individual}}

In this section, we remark on anything that is notable about the
FSPL events from 2016-2018.  For notes on the 2019 events, see 
\citet{kb192073}.

The majority of these notes concern the fact that 10 of the 17 FSPL
events that are reported here were previously published or (in one case)
is the subject of work in preparation.  By contrast, only one of the
13 events analyzed by \citet{kb192073} from 2019 had been the subject
of a previous publication, namely the FFP OGLE-2019-BLG-0551 \citep{ob190551}.
For each of the 9 published events, we compare the best value of $\theta_\e$
reported 
in these papers (first) to the one we report here (second), both in $\muas$:

Two of these events, 
{\bf OGLE-2016-BLG-1540} \citep{ob161540} (9.2 vs.\ 8.8) and
{\bf KMT-2017-BLG-2820} \citep{kb172820} (5.9 vs.\ 5.9)
are FFPs, with the latter having been discovered as part of the 
program being reported here.

Three were published as BD candidates as inferred from 
their relatively small Einstein radii:
{\bf OGLE-2017-BLG-0560} \citep{ob161540} (39 vs.\ 35),
{\bf MOA-2017-BLG-147} \citep{mb19256} (51 vs.\ 46), and
{\bf MOA-2017-BLG-241} \citep{mb19256} (28 vs.\ 26).

Four of these events were the subject of previous publications
because they are FSPL events for which it was possible to derive microlens
parallaxes based on {\it Spitzer} data.
{\it Spitzer} carried out a large, 6-year program whose principal goal
was to measure the microlens parallaxes, and thereby the masses of and
distances to, planetary-system microlenses \citep{yee15}.  However,
{\it Spitzer} microlens parallaxes also yield masses and distances
for FSPL events, whether the finite source effects are observed from
the ground or {\it Spitzer} \citep{ob150763}.  The four published events
are:
{\bf OGLE-2016-BLG-1045} \citep{ob161045} (244 vs.\ 249),
{\bf OGLE-2017-BLG-0896} \citep{ob170896} (140 vs.\ 140),
{\bf OGLE-2017-BLG-1186} \citep{ob171186} (94 vs.\ 94), and
{\bf OGLE-2017-BLG-1254} \citep{ob171254} (207 vs.\ 224).

The reason that the $\theta_\e$ values are identical for three events
(KMT-2017-BLG-2820, OGLE-2017-BLG-0896, OGLE-2017-BLG-1186)
is that we adopted the published parameters after finding that we
recovered the event using our standard procedures.

In addition, S.\ Tsirulik, J.C.\ Yee et al., (in 
preparation) applied this technique to {\bf OGLE-2018-BLG-0705}
and found $D_L = 2.2\pm 0.1\,\kpc$ and 
$M=0.74\,\pm 0.05\, M_\odot$.  The event is projected against the cluster
NGC 6544, and the proper motion measured by {\it Gaia} of the blended
light (almost certainly the lens) is consistent with that of the cluster.
Hence, it is very likely that the lens is a cluster member, having
$2\,\sigma$ tension with the literature-based distance 
to NGC 6544 ($2.6\pm 0.3\,\kpc$).

{\bf OGLE-2017-BLG-0905}:  This event is by far the largest outlier on
the CMD, lying roughly 2 mag above the bulge giant branch (highest point in
Figure~\ref{fig:cmdall}).  This ``problem'' would be ameliorated (but not
completely solved) if the true color were substantially redder than the
one that we measured.  First, however, we have made separate color measurements
from KMTC and KMTS data, and these differ by only 0.04 mag.  Second, the
color (and the magnitude) that are derived from the light curve are
nearly identical to those of the baseline object, as would be expected
for such a bright source.  This is true for both KMTC and KMTS.
Another possible explanation for the extreme brightness of the source
(for its color) is that it is actually a foreground disk star.  However,
the {\it Gaia} proper motion $\bmu_S(E,N) = (-5.3\pm 0.3,-7.5\pm 0.2)\masyr$, 
is  only $\sim 3\,\masyr$ from the centroid of the bulge distribution.  On
the other hand, this $\bmu_S$ would be unusual for a disk star, unless
it were very nearby, which is disfavored by the {\it Gaia}
parallax measurement $\pi_S = 0.13\pm 0.27\,\mas$.  Thus, the source
is most likely a relatively rare post-AGB star.

Although $\bmu_S$ is only $1\,\sigma$ from the bulge mean, this deviation
happens to be close to the direction of anti-rotation, so that 
$\mu_S=9.2\,\masyr$.  This makes our otherwise slightly puzzling
measurement of $\mu_\rel=12.1\,\masyr$ (the second largest in our sample),
more understandable.

Finally, we note that most of the $I$-band points in and near the
peak of the event were saturated.  However, the event parameters
were easily recovered from the $V$-band light curve.  Moreover,
a significant minority of $I$-band points were not saturated due
to a combination of ``poor'' seeing and/or favorable placement of the
source near the pixel corners.  These ``salvaged'' $I$-band points
yielded consistent results with the $V$-band analysis.

Two other events,
{\bf MOA-2016-BLG-258} and
{\bf OGLE-2017-BLG-1186},
were also saturated at peak.
In the first case, the saturation was mild and intermittent, and
saturation did not affect KMTA at all.  Hence, 
no special measures were required to fit the light curve other than
removing a few saturated points.  For the second case, which was already
mentioned above as a published event, the peak region was very densely
covered in the $V$ band due to a combination of the long source self-crossing
time $t_*=\rho t_\e = 3.8\,$days and the high $V$-band cadence
$\Gamma = 0.4\,{\rm hr}^{-1}$.  Hence, there was no difficulty in modeling the
event even though all of the peak $I$-band data had to be excluded.

{\bf OGLE-2017-BLG-0084}:  This event has a relatively large color
error, $\sigma(V-I)_S=0.14$, in a measurement that yields a somewhat 
unexpectedly red source, $(V-I)_{S,0} = 1.33\pm 0.14$, compared to
$(V-I)_{S,0} \sim 1.06\pm 0.10$ for lower red-giant branch stars of 
its apparent magnitude and in its angular proximity.  
The relatively poor measurement
is due to the event peaking at the beginning of the season, when
the nightly visibility was brief, so few $V$-band points were taken
when the $V_s\sim 23.5$ source was highly magnified.  This color is
very similar to that of the baseline object, whose error is also large.
We accept the color measurement at face value.  Nevertheless, we note
that if we had imposed the lower giant-branch color, then the estimates
of $\theta_\e$ and $\mu_\rel$ would have been reduced by a factor 0.83 from
264 to 211 mas, and from 1.71 to $1.42\,\masyr$, respectively.  We do not
impose this prior because it would be inappropriate to do so without
also taking account of the reduced probability of detection, which in
this low-$\mu_\rel$ regime scales roughly as $\theta_\e\mu_\rel^3$, i.e.,
implying a combined factor of 0.47.  The net effect of applying
both priors would be too small to warrant introducing this level of
complexity.

Three events met the initial selection criteria and also yielded 
excellent $\rho$ measurements, but were nevertheless excluded from
the final sample because the sources either are not giants or could not
be properly characterized.  As a matter of due diligence, we document
these decisions.

{\bf KMT-2017-BLG-1725}:  The high magnification of this event enables
a very precise measurement of the $(V-I)$ color, 
despite high extinction, $A_I=3.84$,
revealing that the source lies $\Delta[(V-I),I]=(-1.10,+1.03)$ blueward and
fainter than the clump.  These offsets are inconsistent with a bulge
giant, and indeed with any bulge source, other than from very rare
populations, such as extreme blue horizontal branch stars.  Most likely,
the source lies far in the foreground along this $b=-1.1$ line of
sight, which implies that it
would be probing a very different population of lenses compared to most
of our sources.  In addition, due to the source's unknown extinction,
we cannot reliably estimate $\theta_*$ (so $\theta_\e$) which is the basic
point of the entire project.  We therefore exclude this event from the sample.

{\bf KMT-2018-BLG-1993}: This event lies far out on the near side of the
bar and very close to the Galactic plane $(l,b)=(+6.06,+0.33)$.
The KMT event page lists the extinction as $A_I=5.60$ (derived by 
approximating $A_I = 7\,A_K$, where $A_K$ is from \citealt{gonzalez12}).
The KMT $V/I$ CMD confirms that the field is too extincted to determine
the source properties in these bands.  As we usually do in such cases,
we try to clarify the nature of the source by incorporating
VVV survey data \citep{vvvcat}.  We match KMT $I$-band pyDIA
photometry to $K$-band VVV to
construct an $[(I-K),I]$ CMD.  We find that the clump is at 
$I_{\rm cl,pyDIA}=19.90$,
which combined with $I_{\rm cl,0}=14.27$ from \citet{nataf13}, would yield
$A_I\sim 5.6$ (taking account of the fact that pyDIA typically differs from
standard Cousins $I$ by $\la 0.2\,$ mag whenever it can be calibrated).  
From the light curve model, we obtain $I_{s,\rm pyDIA}=21.79\pm 0.10$. implying
that the source lies $1.89\pm 0.10$ below the clump.  In principle, this
would be consistent with it being a lower giant-branch star in the bar.
We then check the nearest VVV entry and find that it is
$\Delta[(J-K),K] = (-0.40,1.51)$ bluer and fainter than the clump.
This ``catalog entry'' cannot be directly identified with the
microlensed source (with $\Delta I=1.89$) because if it were,
then it would be $\Delta (I-K)=+0.48$ redder than the clump.
Therefore, the VVV entry must contain a blend that contributes
substantial light that is much bluer than the source.  In principle,
the source might then still be a giant at the base of the giant branch,
and, for example, the blend could be the lens, lying far in the foreground
and therefore blue.  Unfortunately, we have no corroborating evidence
that this is the case, and the source could be a subgiant lying much closer
along the line of sight.  The plausibility of this alternate scenario
is augmented by the source's near-plane position, $b=+0.33$.  That is,
the line of sight is never more than 50 pc from the plane, so without
color-magnitude information, the source could be anywhere along this
line of sight.  Hence, we exclude this event.

{\bf KMT-2016-BLG-2608}:  Guided in part by experience with the above
two events, we carefully reviewed all the FSPL events, including those
from 2019 that were analyzed by \citet{kb192073}, in order to identify
problematic events.  This investigation led to the exclusion of one
further FSPL event: KMT-2016-BLG-2608.   This event was the 20th ``new event''
found in our special giant-source search, beyond the 2588 events found
by EventFinder for 2016.  Following the conventions established by
\citet{ob161928} and \citet{kb172820}, we assign it a discovery
sequence number $2588+20=2608$.  Due to very heavy extinction, $A_I=5.56$,
the $(V-I)$ color could not be measured despite relatively high magnification
$A_\max \sim 40$.  Like KMT-2018-BLG-1993, KMT-2016-BLG-2608 lies very
close to the plane, $b=-0.27$, but in contrast to that event, the source
is much brighter than the clump, i.e., by $\Delta I=-1.38$ mag.
In itself, this makes it a very plausible giant-source candidate.
However, we are unable to obtain minimal confirmation of this assessment
from VVV data.  The nearest catalog entry has $K=12.58$ (compared to
$K_{\rm cl}=13.80$).  If we identify this entry with the source,
then it is $\Delta (I-K)\sim -0.16$ bluer than the clump.  This would
be slightly unusual for such a bright star, and we note in particular that
this is not a problem that could be resolved by assuming that the VVV
entry contains blended light: this would only make the source bluer.
These concerns are amplified by the VVV IR color information: there is
only one other band with a flux measurement, $H$, and this yields
$(H-K)_S=-0.04$, compared to $(H-K)\sim 0.8$ for typical bulge giants
in this field.  Because $(H-K)_0$ is strictly positive for all normal
stars, the only plausible explanation for this IR color measurement
(other than measurement error) is that the source lies well in the foreground.
This explanation is also consistent with the very low Galactic latitude
of this line of sight, $b=-0.27$.  In principle, as just mentioned,
it is possible that the $H$-band measurement is incorrect, but the
giant-source scenario requires that this error be about 0.8 mag.
While possible in principle, the balance of evidence favors a foreground
source, so we eliminate this PSPL event.

Finally, we note that according to the above-mentioned naming conventions,
the KMT counterparts to OGLE-2016-BLG-0245 and OGLE-2017-BLG-0560, which
were both re-discovered in the special giant-source searches, are assigned
KMT names KMT-2016-BLG-2627 and KMT-2017-BLG-2830, respectively.  
In contrast to KMT-2016-BLG-2608, these names have no lasting significance
because the official (i.e., discovery) name is from OGLE.  However, these
names are needed in the context of the current paper to maintain
homogeneous conventions.

Nevertheless, this purely formal naming issue does relate to
an important, substantive question: within the sample of 30 FSPL events,
exactly four were not discovered by EventFinder:
OGLE-2019-BLG-0551 $(\rho=4.5)$, 
OGLE-2016-BLG-1540 $(\rho=1.6)$,
KMT-2017-BLG-2820 $(\rho=1.2)$, and
OGLE-2017-BLG-0560 $(\rho=0.9)$.
All four were discovered in the special
giant-source search, but the first has a ``normal'' KMT name because
it had already been discovered by AlertFinder.  This sample of
EventFinder ``failures'' all have high $\rho$.  Indeed, of the 5 FSPL
events with $\rho>0.6$, only KMT-2019-BLG-2073 $(\rho=1.2)$ was discovered
by EventFinder.  Recall that the original motivation for the special
giant-source search was that EventFinder had failed to discover
OGLE-2019-BLG-0551, likely (it was thought) due to its deviations
from the standard form of \citet{pac86} and \citet{gould96} profiles.
By contrast, AlertFinder simply looks for rising light curves,
but this system was not operating (or not fully operating) in 2016-2018.
Hence, the discovery of KMT-2017-BLG-2820 and the tight overlap between
EventFinder ``failures'' and high-$\rho$ events together 
show that this innovation was both necessary and (at least mostly) sufficient.

\acknowledgments

This research has made use of the KMTNet system operated by the Korea
Astronomy and Space Science Institute (KASI) and the data were obtained at
three host sites of CTIO in Chile, SAAO in South Africa, and SSO in
Australia.
Work by C.H. was supported by the grant (2017R1A4A101517) of
National Research Foundation of Korea.
%



\end{document}